\documentclass[5p]{elsarticle}
\usepackage{hyperref}
\usepackage{xcolor}
\journal{}

\begin{document}
\begin{frontmatter}
\title{Spatial and temporal structure of EAS reflected Cherenkov light signal}
\author[address1]{R.A.~Antonov}
\author[address1]{E.A.~Bonvech}
\author[address1]{D.V.~Chernov}
\author[address1]{T.A.~Dzhatdoev\corref{correspondingauthor1}}
\cortext[correspondingauthor1]{Corresponding author}
\ead{timur1606@gmail.com}
\author[address2,address1]{V.I.~Galkin}
\author[address2,address1]{D.A.~Podgrudkov}
\author[address1]{T.M.~Roganova}
\address[address1]{Federal State Budget Educational Institution of Higher Education, M.V. Lomonosov Moscow State University, Skobeltsyn Institute of Nuclear Physics (SINP MSU), 1(2), Leninskie gory, GSP-1, 119991 Moscow, Russia}
\address[address2]{Federal State Budget Educational Institution of Higher Education, M.V. Lomonosov Moscow State University, Department of Physics, 1(2), Leninskie gory, GSP-1, 119991 Moscow, Russia}

\begin{abstract}
A compact device lifted over the ground surface might be used to observe optical radiation of extensive air showers (EAS). Here we consider spatial and temporal characteristics of Vavilov-Cherenkov radiation (``Cherenkov light'') reflected from the snow surface of Lake Baikal, as registered by the \mbox{SPHERE-2} detector. We perform detailed full direct Monte Carlo simulations of EAS development and present a dedicated highly modular code intended for detector response simulations. Detector response properties are illustrated by example of several model EAS events. The instrumental acceptance of the \mbox{SPHERE-2} detector was calculated for a range of observation conditions. We introduce the concept of ``composite model quantities'', calculated for detector responses averaged over photoelectron count fluctuations, but retaining EAS development fluctuations. The distortions of EAS Cherenkov light lateral distribution function (LDF) introduced by the \mbox{SPHERE-2} telescope are understood by comparing composite model LDF with the corresponding function as would be recorded by an ideal detector situated at the ground surface. We show that the uncertainty of snow optical properties does not change our conclusions, and, moreover, that the expected performance of the SPHERE experiment in the task of cosmic ray mass composition study in the energy region $\sim$10 PeV is comparable with other contemporary experiments. Finally, we compare the reflected Cherenkov light method with other experimental techniques and briefly discuss its prospects. 
\end{abstract}
\end{frontmatter}

\section{Introduction}

Direct studies of high-energy cosmic rays (CR) with balloon and satellite experiments are available only below $E= 10^{15}$ eV = 1 PeV \cite{asa98,der05,pan17,yoo17}. Nearly all knowledge about CR above this energy is being obtained with indirect methods, i.e. through observation of extensive air showers (EAS) --- cascades of particles initiated in the atmosphere by primary nuclei.

The majority of EAS experiments use a grid of coherently working detectors distributed over the Earth's surface that are sensitive to various EAS components. Many EAS arrays were designed to study charged particles, e.g.\ EAS-TOP \cite{agl89}, Tibet-III \cite{ame08},  GAMMA \cite{gar08}, KASCADE-Grande \cite{ape10, ape12, ant05}, IceTop \cite{abb13, icetop13}, the surface detectors of the Telescope Array (TA) \cite{abu12}, Yakutsk \cite{iva13} and Pierre Auger Observatory (PAO) \cite{aab15} experiments. These surface detectors are usually also sensitive to high-energy $\gamma$-rays that accompany EAS electrons and positrons (hereafter simply ``electrons''). Experiments with optical Vavilov-Cherenkov radiation (hereafter simply ``Cheren\-kov light'') include CASA-BLANCA \cite{fow01}, BASJE \cite{tok08}, TACT~\cite{ant95} and the optical part of the Yakutsk array \cite{iva09}. EAS fluorescent light was observed with HiRes \cite{abb08} and the optical detectors of TA \cite{tok12,abb16} TALE~\cite{tale15} and PAO \cite{aab15}. EAS radio emission is also studied in various experiments, such as LOPES \cite{ape14b}, PAO \cite{abr12}, and LORA \cite{tho16a}.

The size of ground-based EAS detector arrays and their com\-ple\-xi\-ty is growing with time, and so does the difficulty of their deployment, calibration, and operation. Indeed, in order to work as an ensemble, the detectors must be distributed over a large area, sometimes hundreds \cite{abu12} or thousands \cite{aab15} square kilometers, and, moreover, they must be constantly power-supplied, be able to transfer information obtained with them, and kept fairly well time-synchronized.

A long time ago it was pointed out that a compact device lifted over the snow-covered Earth surface is able to register reflected Cherenkov light \cite{chu74}. This method is free from the above-mentioned difficulties of ground-based experiments. Moreover, a detector looking down at reflected Cherenkov light typically has a quasi-continious spatial sensitivity, i.e. it can observe light emitted from a substantial fraction of the snow-covered surface. This allows to resolve a sharp peak in the EAS Cherenkov light lateral distribution function (LDF) that is typically present near the axis for EAS with moderate primary zenith angle ($\theta<$30-40$^{\circ}$). In contrast, ground-based EAS arrays, as a rule, can not probe Cherenkov light properties near the EAS axis due to limited sensitive area of their detectors (usually not exceeding several square meters).

A historical review of this method is available in~\cite{ant15a}. In fact, such an approach is also applicable to the registration of reflected EAS radio emission \cite{sch15}. Fluorescent light, due to its isotropic angular distribution, can be observed directly by a balloon-borne \cite{ant06,ant16b,ant17} or satellite \cite{tak95} experiment. The JEM-EUSO detector that will be able to observe both fluorescent and reflected Cherenkov light emission of EAS from space is under development \cite{tak09, adam15a,adam15b,adam15c,abd17}.

\begin{figure*}[tb]
\begin{minipage}{0.47\textwidth}
\centering
\includegraphics[width=18pc]{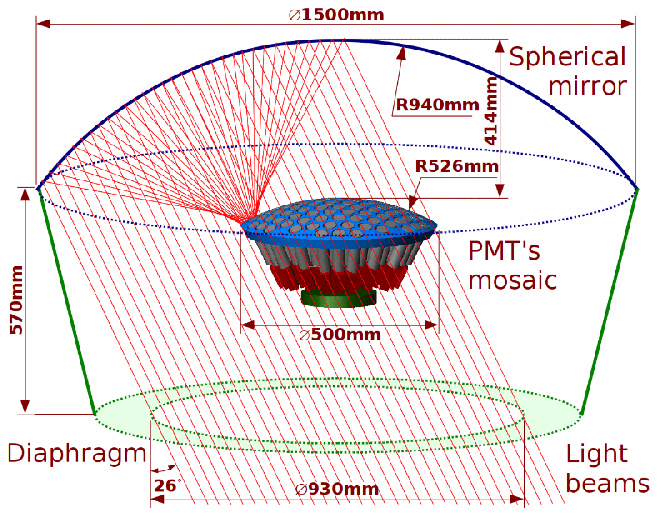}\hspace{2pc}%
\end{minipage}
\hfill
\begin{minipage}{0.47\textwidth}
\centering
\includegraphics[width=18pc]{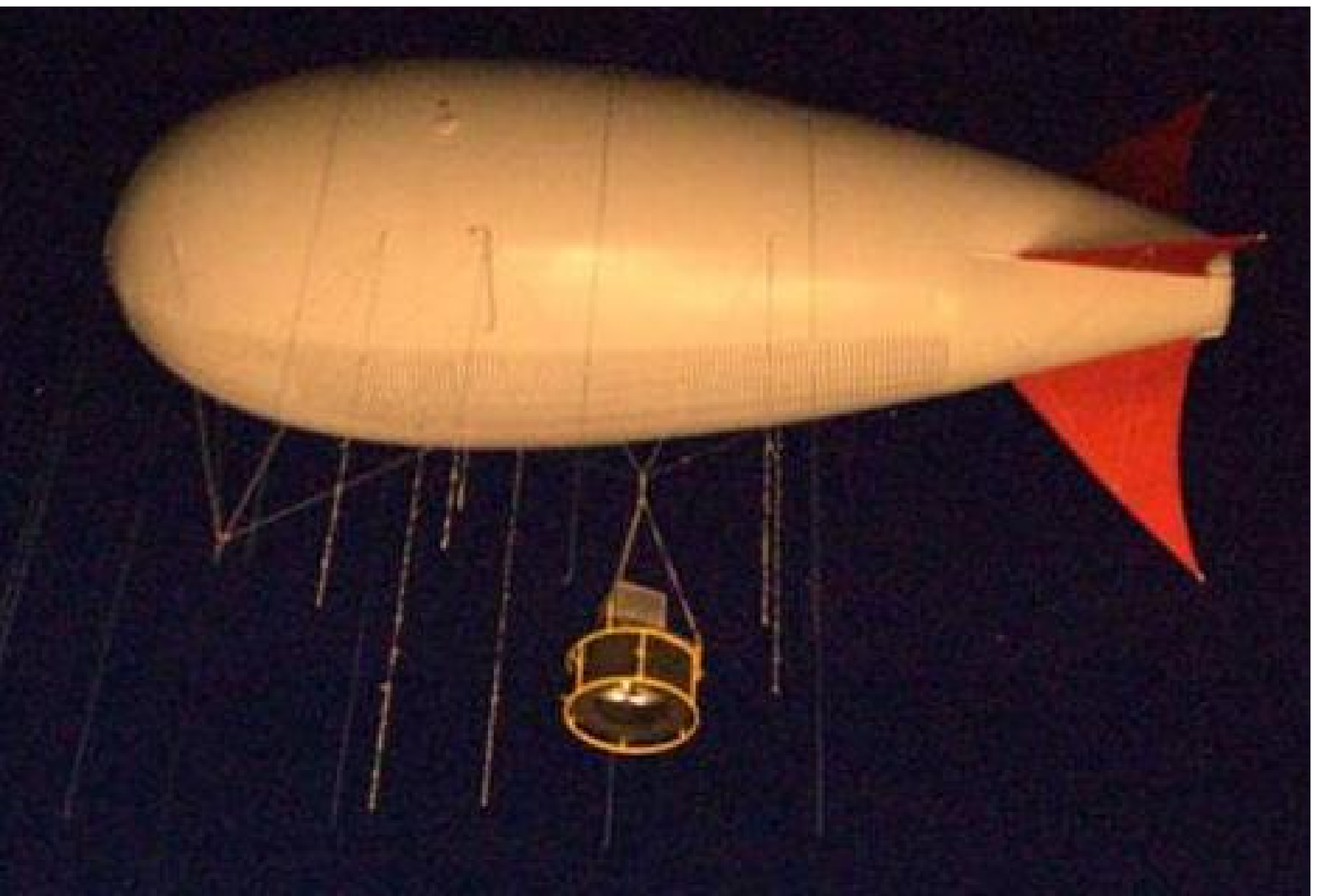}\hspace{2pc}%
\end{minipage}
\vfill
\begin{minipage}[b]{0.47\textwidth}
\caption{The SPHERE-2 telescope optical scheme.}
\label{fig1}
\end{minipage}
\hfill
\begin{minipage}[b]{0.47\textwidth}
\caption{The SPHERE-2 telescope carried by the BAPA tethered balloon.}
\label{fig2}
\end{minipage}
\end{figure*}

The simplest and most cost-ef\-fec\-ti\-ve experimental technique of such kind appears to be the registration of reflected Cherenkov light with a telescope lifted above the ground surface at modest altitude $H<$ 1--3 km. Indeed, a much higher observation altitude would require the involvement of an expensive stratospheric balloon or other aircraft, not to mention satellite. At low altitudes ($H<$3 km) reflected Cherenkov signal as a rule is much brighter than direct or reflected fluorescent signal, due to higher light yield and the fact that Cherenkov light from EAS in the atmosphere has very sharp directional pattern, thus accumulating over the EAS development path.

The \mbox{SPHERE-2} telescope is currently the most advanced detector employing the reflected Cherenkov light method \cite{ant15a} (for a brief review  see \cite{ant15b}). In \cite{ant13a,ant13b} the possibility to reconstruct the all-nuclei spectrum of primary CR using the observations performed with this detector was demonstrated. Moreover, a method for an event-by-event study of the CR mass composition in the energy range 10-100 PeV using the LDF steepness was devised \cite{ant15c}.

Robust measurement of the all-nuclei spectrum and nuclear composition by means of reflected Cherenkov light requires clear understanding of basic features of the signal, as well as reliable models of EAS development and detailed detector response simulations. In this paper, for the first time, we present a detailed calculation of temporal and spatial structure of the reflected Cherenkov light signal, including effects introduced by the \mbox{SPHERE-2} detector. In Sect.~\ref{sect:detector} the \mbox{SPHERE-2} detector is briefly described, in Sect.~\ref{sect:ground} the simulation of EAS Cherenkov light spatial and temporal distributions at the ground level is considered. Sect.~\ref{sect:response} deals with the reflection from the snow cover and the response of optics and electronics of the \mbox{SPHERE-2} detector. In Sect.~\ref{sect:acceptance} we describe the simulation of the \mbox{SPHERE-2} telescope trigger response and instrumental acceptance. Notwithstanding the high level of fluctuations caused by low number of photons registered by the detector, a highly modular structure of our code allows to study the distortions caused by the detector's non-ideality. This is achieved by averaging over many responses generated as replicas from one certain model EAS (Sect.~\ref{sect:composite}). Sect.~\ref{sect:snow} deals with measured and simulated optical properties of realistic snow covers. In this Section we show that contemporary understanding of snow optical properties is sufficient to reconstruct the distribution of the EAS LDF steepness parameter without an appreciable distortion. In Sect.~\ref{sect:composition} we discuss the performance of the SPHERE detector for the specific task of measuring the spectrum of CR light component (i.e. the combined spectrum of proton and Helium nuclei). In Sect.~\ref{sect:discussion} we briefly compare the reflected Cherenkov light technique with other approaches used to study the CR spectrum and composition and discuss the prospects of the reflected Cherenkov light method. Finally, we draw out our conclusions in Sect.~\ref{sect:conclusions}.

\section{The SPHERE-2 detector \label{sect:detector}}

\begin{table}[b]
\centering
\begin{tabular}{l l l}
\hline
\textbf{Parameter} & \textbf{ Value}\\
\hline
Detector field-of-view & 52$^\circ$ \\
Mirror's area \\(with account of the shadow \\ from the mosaic) & 0.48 m$^2$ \\
Geometric area \\(without account of the shadow \\ from the mosaic) & 0.68 m$^2$ \\
Optical point-spread at the mosaic center \\(FWHM diameter) & 34 mm \\
Optical point-spread at the mosaic edge \\(main axes FWHMs) & 34x23 mm \\
Time resolution (ADC) & 12.5 ns \\
Time resolution \\(phase shift between channels) & $\sim$3 ns \\
\hline
\end{tabular}
\caption{Some basic parameters of the SPHERE-2 detector \label{tab1}}
\end{table}

The SPHERE-2 instrument was designed to observe Cherenkov light of EAS, reflected from the surface of Lake Baikal (south-east Siberia, Russian Federation). Let us briefly recall some information about the \mbox{SPHERE-2} apparatus essential to understand our simulation. More details on the detector's optics and electronics could be found in \cite{ant15a}. 

The optical scheme of the detector is shown in Fig.~\ref{fig1}. It consists of a spherical mirror, a mosaic of photomultipliers (PMTs) and an aperture diaphragm. The mosaic carries 108 PMT-84-3 and one (central) PMT Hamamatsu R3886, included for the calibration purposes. Table~\ref{tab1} shows some parameters of the SPHERE-2 telescope for the 2013 observation run conditions. The typical spectral profile of quantum efficiency $\epsilon(\lambda)$ for Hamamatsu R3886 and PMT FEU-84-3 is shown in Fig.~7 of \cite{ant16}. The relative calibration of the mosaic was performed with 7 light-emitting diodes (LED), situated on the mirror's surface. The calibration method of the \mbox{SPHERE-2} telescope was described in detail in~\cite{ant16}.

\begin{figure*}[bt]
\begin{minipage}{0.47\textwidth}
\centering
\includegraphics[width=18pc]{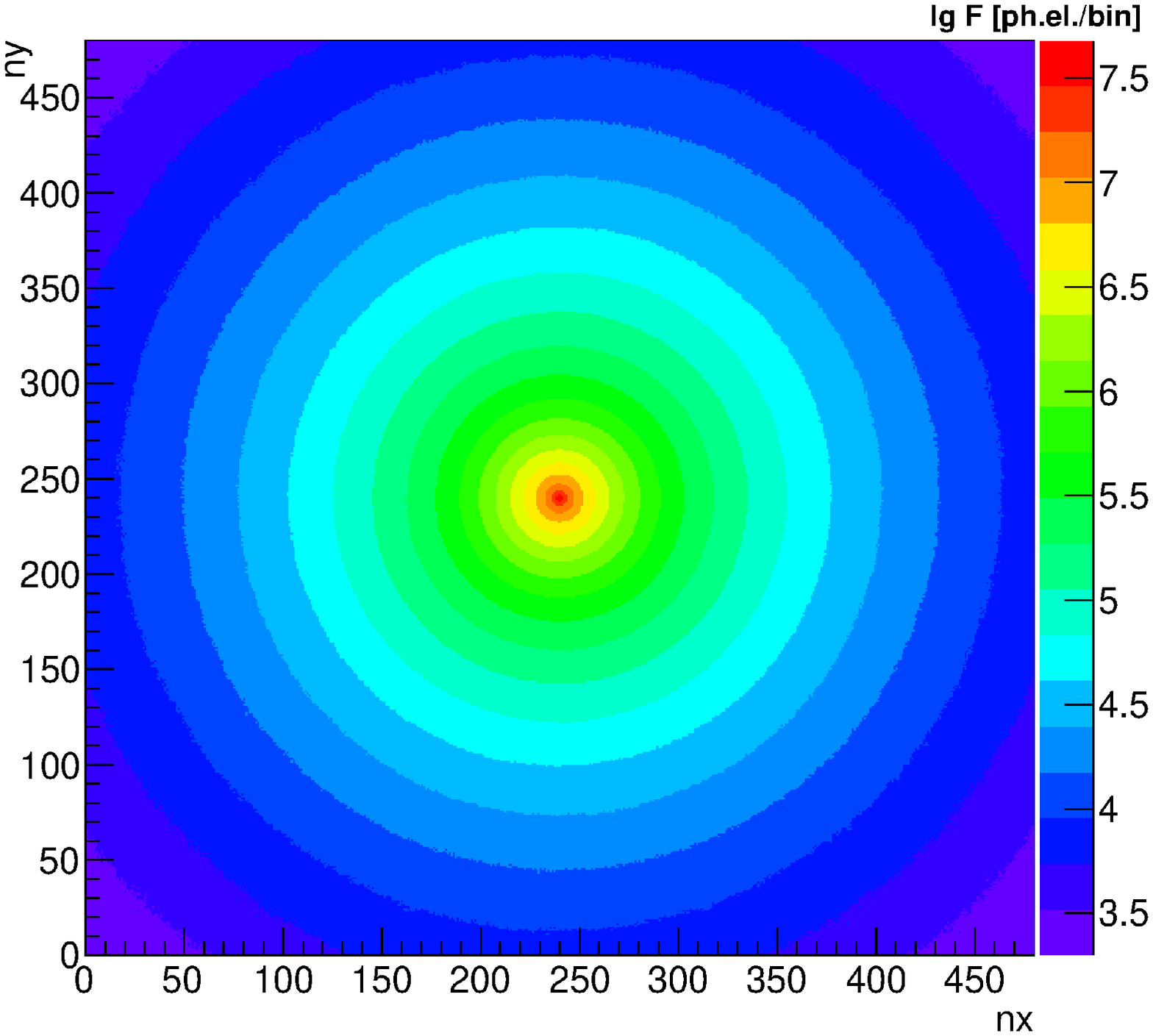}
\caption{An example of Cherenkov light LDF of a shower from primary proton with energy $E$= 10~PeV and $\theta$= 0.20~rad.}
\label{fig3}
\end{minipage}
\hfill
\begin{minipage}{0.47\textwidth}
\centering
\includegraphics[width=18pc]{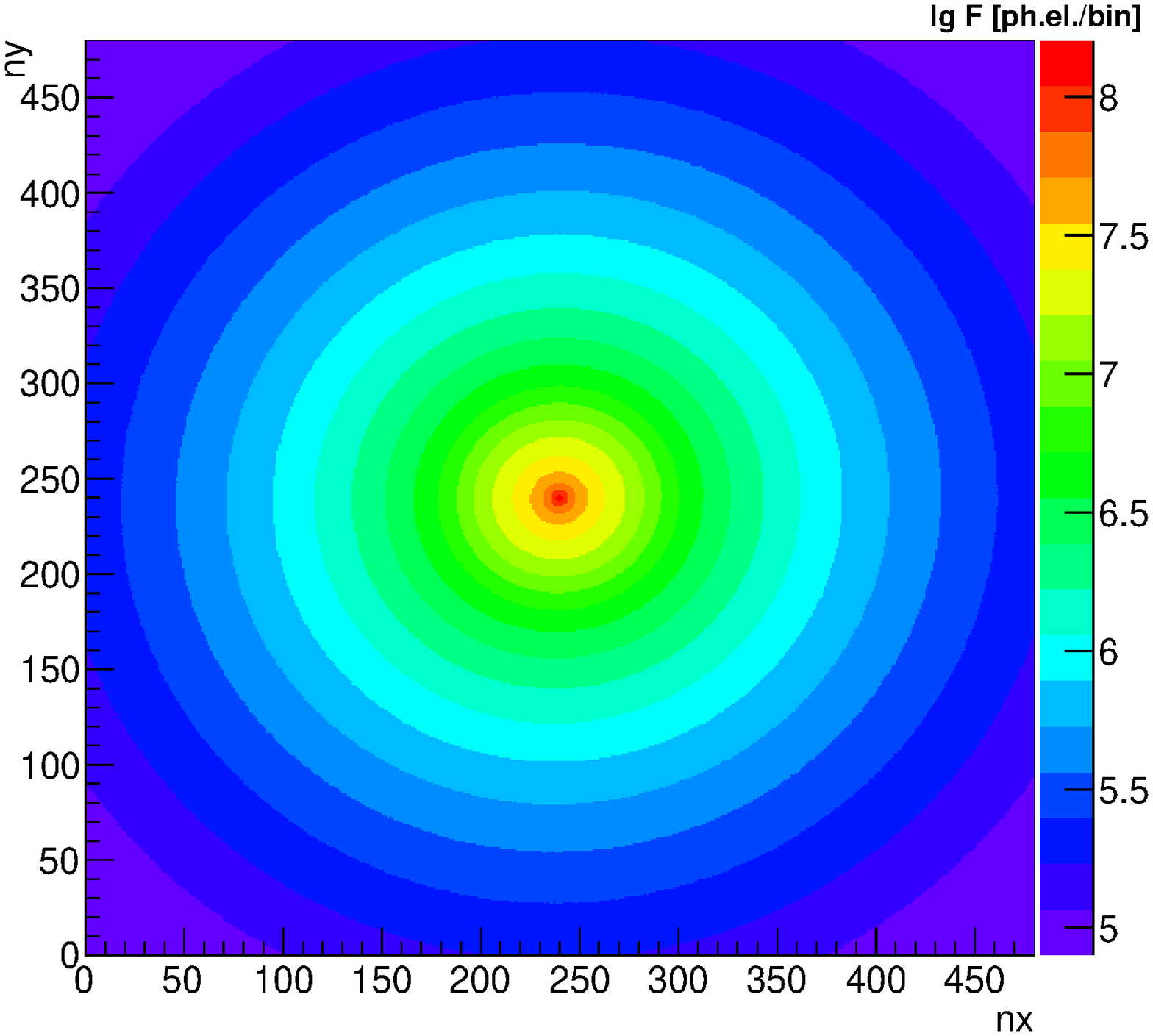}
\caption{The same, as in Fig.~\ref{fig3}, but for $E$= 100~PeV and $\theta$= 0.30~rad.}
\label{fig4}
\end{minipage}

\begin{minipage}{0.47\textwidth}
\centering
\includegraphics[width=18pc]{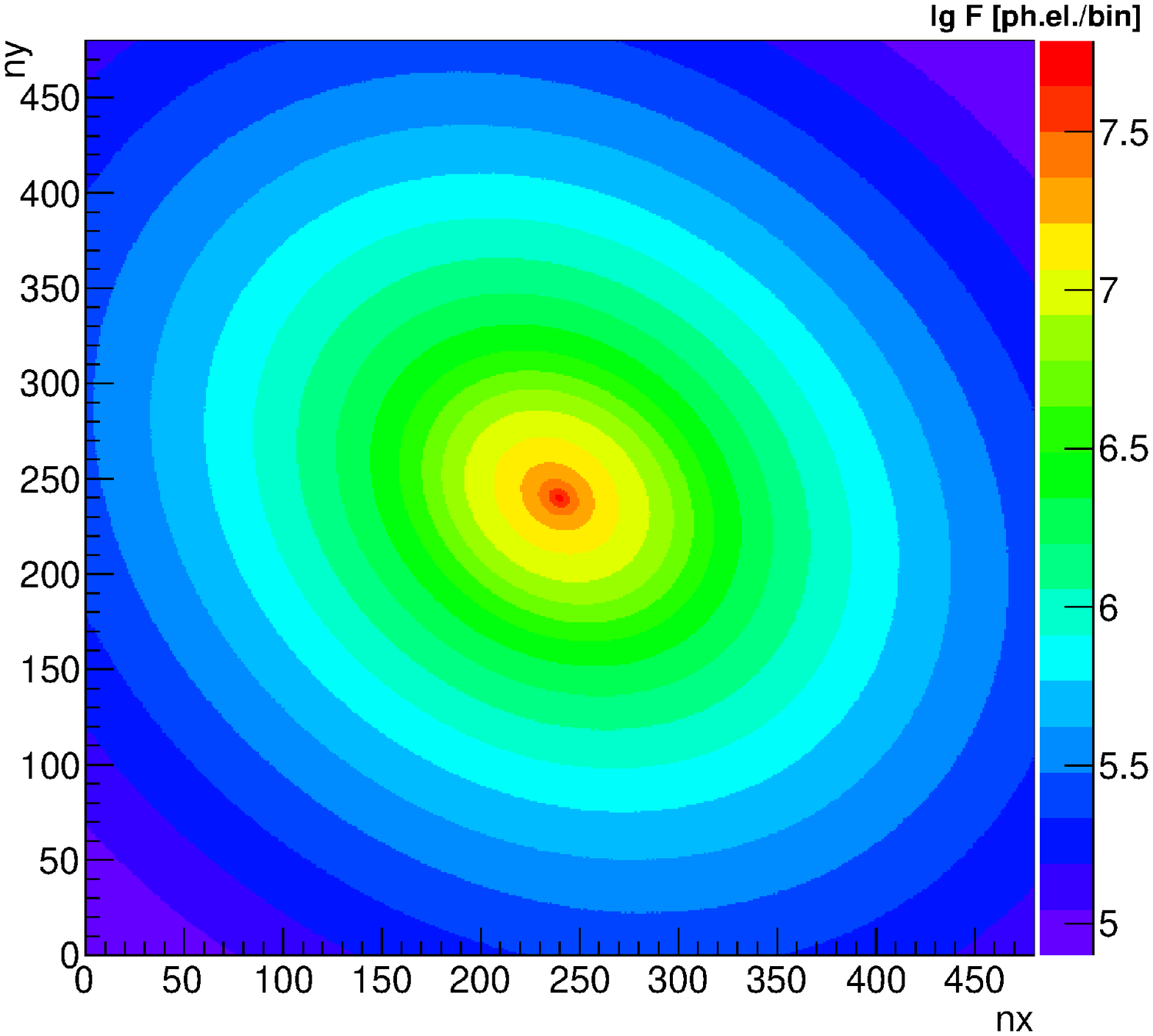}
\end{minipage}
\hfill
\begin{minipage}{0.47\textwidth}
\centering
\includegraphics[width=18pc]{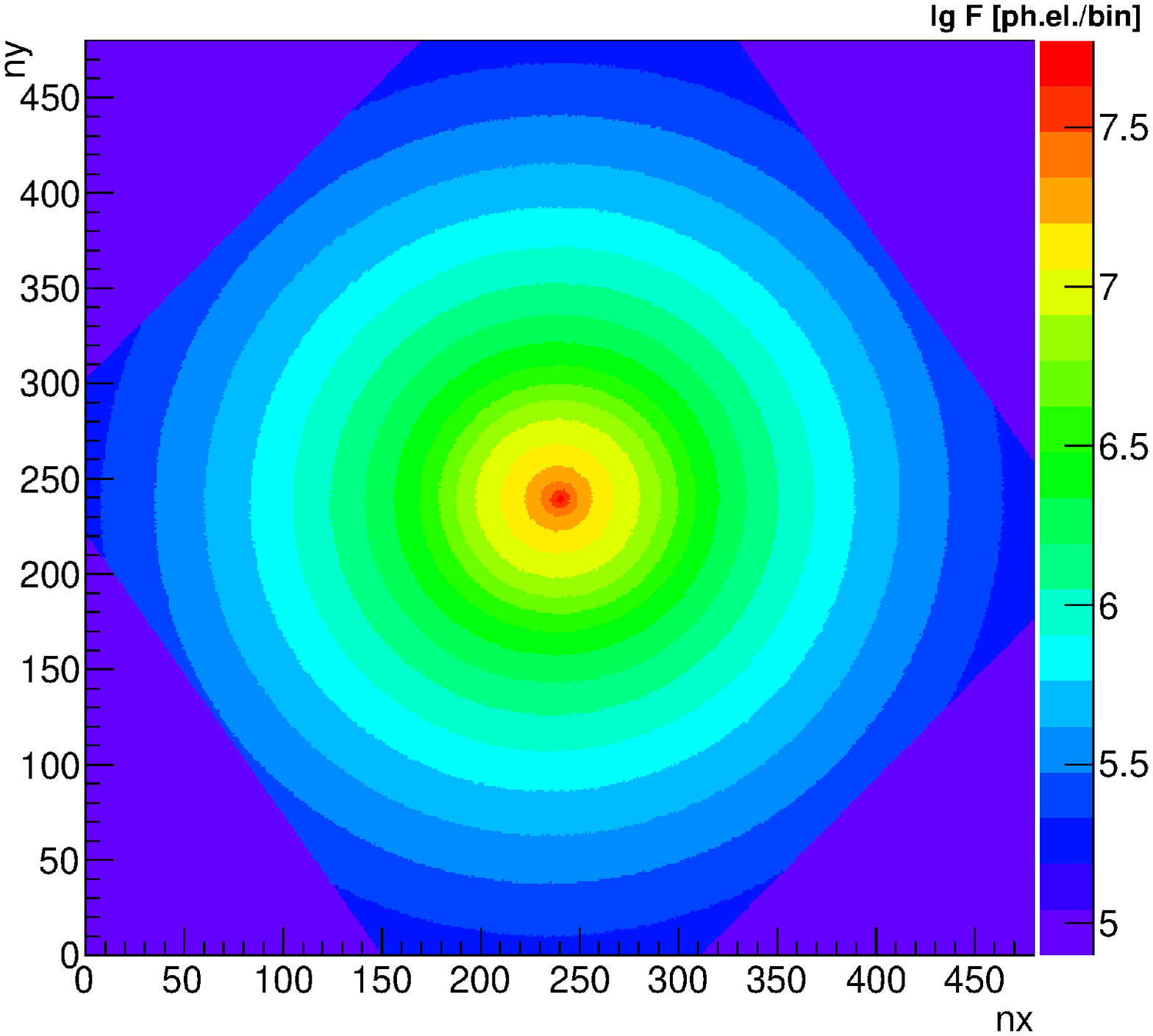}
\end{minipage}
\vfill
\begin{minipage}{0.47\textwidth}
\centering
\caption{The same, as in Fig.~\ref{fig4}, but for $\theta$= 0.63~rad.}
\label{fig5}
\end{minipage}
\hfill
\begin{minipage}{0.47\textwidth}
\centering
\caption{The same, as in Fig.~\ref{fig5}, but after application of the symmetry restoration procedure, explained in the text.}
\label{fig6} 
\end{minipage}
\end{figure*}

The mosaic is provided with readout electronics that records PMT anode pulse shape by 109 analog-to-digital convertors (ADC). For the 2013 run the ADC time sampling step was 12.5 ns. The trigger condition is based on the quantity $S_{4}$ that is defined for each channel as the sum of four ADC measurements separated by the 25~ns time interval each. As well, each channel has an amplitude discriminator (AD) that determines if the $S_{4}$ value exceeds some threshold $S_{thr}$. At the beginning of each flight, when the \mbox{SPHERE-2} detector is already lifted up to the working altitude and ready to start measurements, the threshold values $S_{thr}$ in all channels are adjusted so that the AD rate does not exceed some predefined value (1 $Hz$ for the case of the 2013 run). The trigger condition is satisfied if one of the following statements is true: a) $S_{4}$ exceeds $S_{thr}$ in 3 adjacent channels inside the time window 1 $\mu s$ (the so-called ``local'' condition $L3$) or b) $S_{4}$ exceeds $S_{thr}$ in 5 arbitrary channels inside the same time window (the ``global'' trigger condition $G5$).

A general view of the \mbox{SPHERE-2} telescope suspended under the BAPA (an abbreviation from Russian ``Baikal tethered balloon'') balloon is shown in Fig.~\ref{fig2}. The observation altitude for the 2013 run, measured with the GPS, was in range of 170--710~m (most of the time between 400~m and 700~m). The detector is also provided with an ``inclinometer'' that is able to measure the inclination of the detector's optical axis with respect to the nadir direction.

\section{EAS Cherenkov light signal at the ground level \label{sect:ground}}

Our final goal of studying the primary CR spectrum and mass composition requires a realistic account of EAS development fluctuations, as well as additional fluctuations and distortions introduced by the detector. Now we proceed step-by-step and describe our calculations, starting from Cherenkov light properties at the ground level.

The model of EAS Cherenkov light spatial and temporal properties used in the present work is based on full Monte Carlo (MC) simulations of a large sample of EAS calculated with the CORSIKA 6.500 code \cite{hec98} assuming the \mbox{QGSJET-I} (version 01c) high-energy hadronic model \cite{kal97} and the GHEISHA 2002d low-energy hadronic model \cite{fes85}. While it is known that some parameters of Cherenkov light LDF and pulse shape are dependent on the hadronic model, here we are interested mainly in general properties of EAS Cherenkov light, that are much less model-dependent. Therefore in this paper it is sufficient to use only one hadronic model option \mbox{QGSJET-I}+GHEISHA.

\begin{figure*}[tb]
\begin{minipage}{0.47\textwidth}
\centering
\includegraphics[width=18pc]{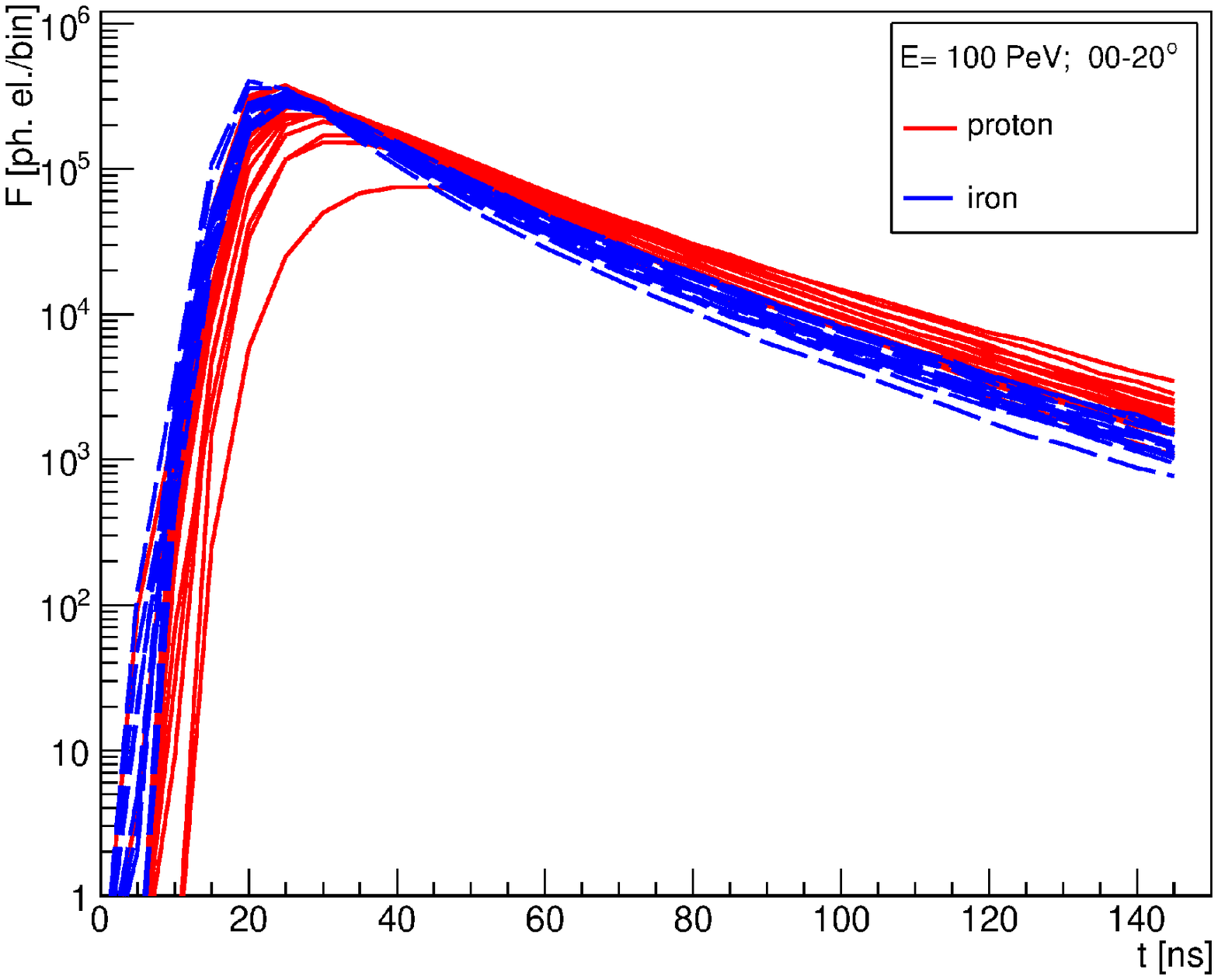}
\end{minipage}
\hfill
\begin{minipage}{0.47\textwidth}
\centering
\includegraphics[width=18pc]{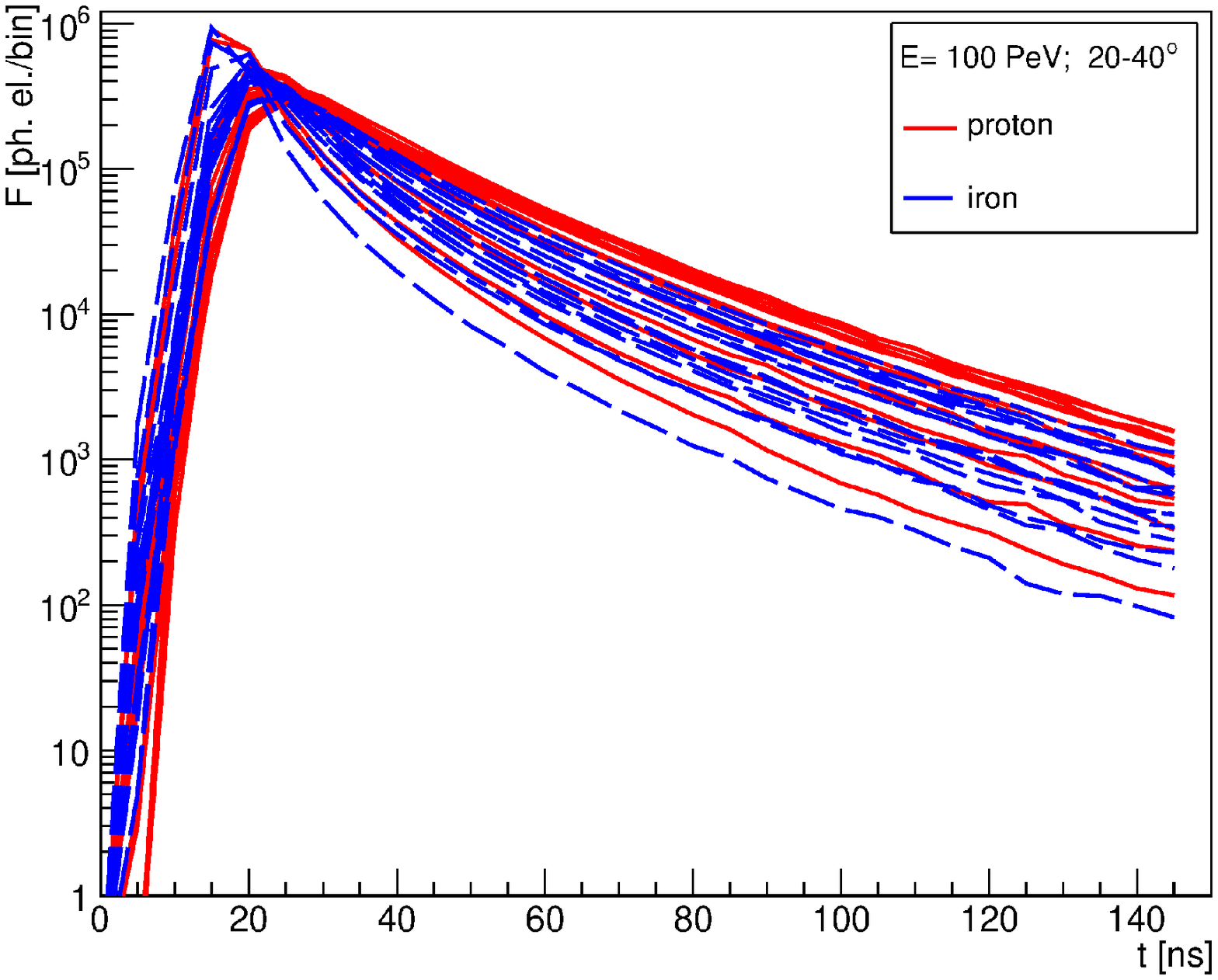}
\end{minipage}
\vfill
\begin{minipage}{0.47\textwidth}
\centering
\caption{Pulse shape for a selection of showers with zenith angle $ \theta<20^{\circ}$. Primary protons --- solid red curves, primary Iron --- dashed blue curves.}
\label{fig7}
\end{minipage}
\hfill
\begin{minipage}{0.47\textwidth}
\centering
\caption{The same, as on Fig.~\ref{fig7}, but for zenith angles $20^{\circ}<\theta<40^{\circ}$.}
\label{fig8}
\end{minipage}
\end{figure*}

The threshold for electrons was set to 19 MeV, slightly below the minimal Cherenkov threshold for these particles in the atmosphere. Cherenkov photon bunch size parameter was set to 1.0 (i.e.\ 1 photon/bunch), and the observation level --- to 455~m a.~s.~l., corresponding to the experimental conditions. The PMT 84-3 quantum efficiency was accounted for so that only a part of photons was propagated to the ground; therefore the Cherenkov light intensity that is reported below is already normalized to the expected number of photoelectrons (ph. el.). Thanks to large number of Cherenkov photons in EAS for the considered primary energy range $E>$1 PeV, this does not introduce any significant additional uncertainty in the simulations. As well, the approximate wavelength-average value of the mirror reflection factor ($K_{Mirr}$=0.9) was accounted for at this step of the simulation.

The result of the simulation for one shower is a three-dimensional histogram $F(nx,ny,nt)$ (where $(nx,ny,nt)$ are numbers of bins along $(x,y,t)$ axis, respectively) with 480 $\times$ 480 spatial bins of extension 2.5 m $\times$ 2.5 m each and 102 time bins. The axis position of all showers is set to the center of the spatial part of the histogram. Time delays are counted from the plane that is normal to the shower's axis (the so-called ``shower's plane''). The first temporal bin serves as a control one to check that the histogram was properly initialized and the arrival time of the first photon with respect to the shower's plane $t_{0}$ was correctly calculated. If any photon arrived before $t_{0}$, it would be recorded in this bin of the histogram. Each of the next 100 bins has temporal width 5 ns; the last one accumulates all photons with arrival time more than 500 ns with respect to the first photon.

Now let us illustrate our simulations with several examples of showers from primary proton. The first shower (hereafter \textit{Shower 1}) has the primary energy $E$ = 10 PeV and the zenith angle of the primary particle $\theta$ = 0.20~rad = 11$^{\circ}$. The second shower (hereafter \textit{Shower 2}) has $E$ = 100 PeV and $\theta$ = 0.30 rad = 17$^{\circ}$. The third EAS (hereafter \textit{Shower 3}) is also for the case of $E$ = 100 PeV, but the larger value of $\theta$ = 0.63 rad = 36$^{\circ}$.

The LDF of \textit{Shower 1}, obtained with direct summation over all time bins, is shown in Fig.~\ref{fig3} (this and most other graphs in the present paper were produced with the ROOT software \cite{roo15}). Every spatial bin in Fig.~\ref{fig3} is 2.5 m wide. The numbers of horizontal and vertical bins are shown near the corresponding axes, so that the full spatial extension of the depicted LDF is 1200~m $\times$ 1200~m. The decimal logarithm of intensity is shown by color. This EAS is nearly vertical, therefore its LDF is almost axially symmetric. The LDF, as is typical for a nearly vertical EAS with sufficiently high energy, has a sharp peak near the axis, thanks to the sharp directional pattern of Cherenkov light in the atmosphere. The LDF of \textit{Shower 2}, that also has a modest value of $\theta$, is shown in Fig.~\ref{fig4}. The intensity of Cherenkov light at the distance from the axis $R\approx$ 150~m is nearly proportional to the shower's energy --- a well known fact \cite{hil82,pat83} that could be used to roughly estimate the energy of the primary particle.

\begin{figure*}[tb]
\begin{minipage}{0.47\textwidth}
\centering
\includegraphics[width=18pc]{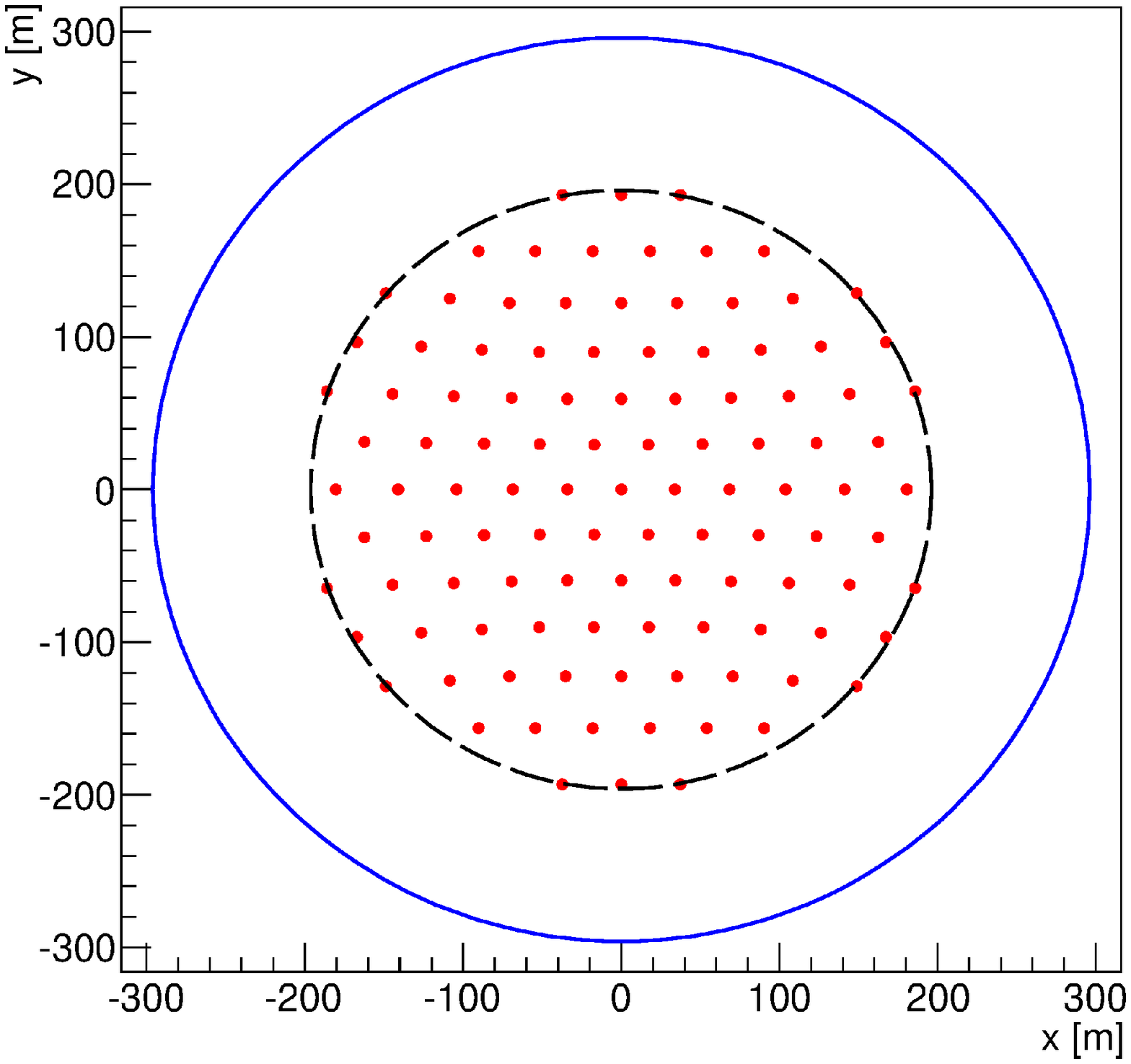}
\caption{Coordinates of points associated with measurement channel centers for the case of $H$= 400 m (small red filled circles) together with the \mbox{SPHERE-2} detector field of view border with the radius $R_{FOV}$ (dashed black circle). $R_{FOV}$+100 m circle is also shown (blue circle).}
\label{fig9}
\end{minipage}
\hfill
\begin{minipage}{0.47\textwidth}
\includegraphics[width=18pc]{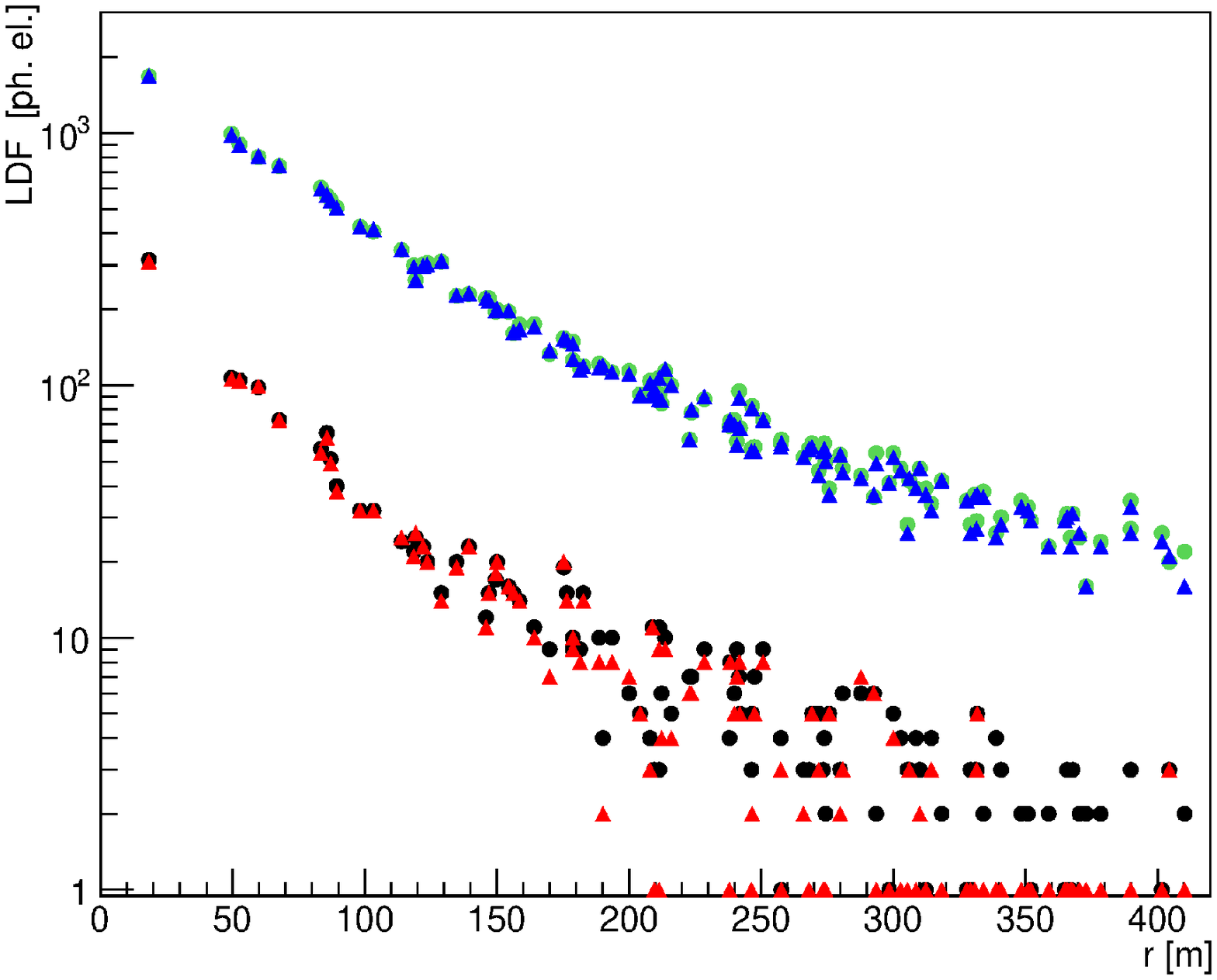}\hspace{2pc}%
\caption{Model LDF before (circles) and after (triangles) digitization procedure for \textit{Shower 1} (lower data) and \textit{Shower 2} (upper data).}
\label{fig10}
\end{minipage}
\end{figure*}

The LDF of \textit{Shower 3} is shown in Fig.~\ref{fig5}. This LDF is less steep than the LDF of \textit{Shower 2}, what is typical for showers with larger zenith angles due to descrease of Cherenkov light production after the shower's maximum. As well, one can see that the equal intensity levels in Fig.~\ref{fig5} are clearly of elliptical shape, again due to sufficiently large zenith angle. This is a well known fact and applies also to the case of LDF of charged particles \cite{sim11}. To restore the azimuthal symmetry of Cherenkov light LDF, we use the following simple procedure: first, the LDF is rotated to make the long axis of the LDF ellipse horizontal and the shower direction coincident with the $x$-axis, and then shrinked along the horizontal axis to the factor $K_{Sh}= cos\,\theta$. The LDF of the \textit{Shower 3} after this ``cylindrical projection'' procedure is shown in Fig.~\ref{fig6}. The azimuthal symmetry is indeed almost restored after the application of the described procedure (however, this may not be the case for low-energy ($E<$1 PeV) or highly inclined ($\theta>$1 rad) showers, see~\cite{hom15}). A part of the spatial histogram that became empty due to the shrinkage of the LDF was filled by the minimal value inside the shrinked part of the histogram. As we are interested in LDF values only well inside the central 600 m, this doesn't affect our results. 

An example of simulated pulse shape for a selection of showers is shown in Figs.~\ref{fig7}--\ref{fig8}. The primary energy $E=100$~PeV, and the distance from the axis $R= 250$~m in all cases. The fluctuations of the pulse shape for primary protons are stronger than for primary Iron, but, nevertheless, the shape is qualitatively similar for most considered showers, irrespectively of the primary nucleus mass.

\section{Detector response simulation \label{sect:response}}

\subsection{Reflection from snow and the detector's optical response \label{ssect:response-optics}}

Spatial and temporal structure of EAS Cherenkov light undergoes certain transformations after scattering by the snow cover and light propagation from the ground to the \mbox{SPHERE-2} telescope; additional distortions and fluctuations are as well introduced by the \mbox{SPHERE-2} detector itself. In this Subsection we account for these effects.

The mean expected number of Cherenkov photons from EAS falling into the diaphragm of the \mbox{SPHERE-2} telescope $N_{D}$ is many orders of magnitude smaller than the full number of Cherenkov light photons reaching the ground level $N_{gr}$. The amount of energy $\delta E$ scattered from the surface element $\delta S$ into an element $\delta \Omega$ of the solid angle is $\delta E= B\,cos\,\theta_{n}\,\delta S\, \delta \Omega$ (\cite{bor70}, p. 181, equation (1)), where $\theta_{n}$ is the angle between the direction to the $\delta \Omega$ element and the normal to the surface, and $B$ is photometric brightness. We note that reflection from snow surface is not mirror-like, but rather has a character of diffuse scattering. For instance, in case of normal incidence of a well-collimated beam, reflected radiation still has a broad angular pattern with shape defined by the dependence of $B$ on $\theta_{n}$.

For a certain radiation spectrum $\delta E$ is proportional to the number of photons $\delta N$ scattered to the same elements of the surface and solid angle. Let us denote as $\delta N_{gr}$ the number of Cherenkov photons, radiated by EAS and falling into the element of surface $\delta S$, and $\delta N_{D}$ --- the number of photons, reflected from the same element of surface and falling into the diaphragm. Assuming that $B$ is independent of $\theta_{n}$ (a good approximation for the case of snow surface \cite{ant15a}) and that the surface is homogeneous, for the case of the \mbox{SPHERE-2} detector observation conditions one can obtain the following relation between $\delta N_{D}$ and $\delta N_{gr}$:
\begin{eqnarray}
\delta N_{D}/\delta N_{gr}= K\cdot K_{1}\cdot K_{2}, \label{eqn1}\\
K_{1}= \frac{R^{2}_{D}}{H^{2}}, \label{eqn2}\\
K_{2}= \frac{H^{2}}{H^{2}+x^{2}+y^{2}} \cdot cos^{2}\theta_{n}. \label{eqn3}
\end{eqnarray}
Here it was assumed that the snow albedo $K$, defined as the ratio of the upward-going flux to the downward-going flux at a particular wavelength $\lambda$ (for instance, see~\cite{war82}), is independent of $\lambda$ and equals to 0.9. The diaphragm's radius is $R_{D}$, the coordinates of the detector are $(0,0,H)$, the coordinates of the reflecting element of the sufrace are $(x,y,0)$, and the telescope's optical axis is vertical. Here we neglected the curvature of the Earth's surface, which is a valid approximation for the case of the altitude $H<1$ km.

The factor $K_{1}$ is by far the main one that determines the $N_{D}$ value provided that $N_{gr}$ is known. The second power of $cos\theta_{n}$ in (\ref{eqn3}) is due to the decrease of the diaphragm's effective area for non-normal incidence. In what follows we use equations (\ref{eqn1})--(\ref{eqn3}) in our calculations. Any small deviation from these equations may be taken into account in the later stages of the simulation. In particular, optical properties of the Lake Baikal snow cover differ from the simplified model described above. The impact of these effects on model detector response events is discussed in Sect.~\ref{sect:snow}.

Technically, the calculation of $N_{D}$ is organised as follows. The first approximation of the $N_{D}$ value was sampled according to the Poisson distribution with parameter $N_{D0}= K\cdot K_{1}\cdot N_{gr}$. Then, the incident coordinates of photons on the snow surface and ``intrinsic'' (i.e. counted from the shower's plane) time delay of each photon $t_{1}$ were simulated with MC technique according to the three-dimensional distribution $F(nx,ny,nt)$. Full time delay of photons arriving at the diaphragm is
\begin{eqnarray}
t= t_{1}+t_{2}+t_{3}, \label{eqn4}\\
t_{2}= (1/c)\cdot(x\cdot sin\theta\ cos\phi +y\cdot sin\theta\ sin\phi), \label{eqn5}\\
t_{3}= (1/c)\cdot(\sqrt{H^{2}+x^{2}+y^{2}}-H), \label{eqn6}
\end{eqnarray}
where $\phi$ is the azimuthal angle of the primary particle and $c$ is speed of light in vacuum. Near the surface of Lake Baikal the refractive index of air $n_{a} \approx 1+2.7\cdot 10^{-4}$ (\cite{hec98}, p. 54) is very close to 1 (and the value of $n_{a}-1$ decreases with increasing altitude), so the difference between speed of light in vacuum and air was neglected.

A part of photons was accepted according to probability $K_{2}$; remaining photons were rejected in order to satisfy equation (\ref{eqn1}). Finally, accepted photons were traced through the optical system of the \mbox{SPHERE-2} detector with a separate Geant4 \cite{ago03} application. A model of the optical system was implemented in the \textit{\textbf{T01Detector}} class that was derived from the \textit{\textbf{G4VUserDetectorCon\-struc\-tion}} class. It consists of a spherical mirror and a non-transparent mosaic. Photons were injected at the position of the diaphragm aperture; some of these photons were absorbed by the mosaic. The mirror reflection factor (that is equal to 0.9), it will be remembered, was already accounted for in Sect.~\ref{sect:ground}. Reflection from PMT glass was taken into account for the case of unpolarized light by rejecting a part of photons defined as:
\begin{eqnarray}
R= \frac{R_{1}+R_{2}}{2}, \label{eqn7}\\
R_{1}= \left(\frac{cos\theta_{i}-n_{g}cos\theta_{t}}{cos\theta_{i}+n_{g}cos\theta_{t}}\right)^{2}, \label{eqn8}\\
R_{2}= \left(\frac{n_{g}cos\theta_{i}-cos\theta_{t}}{n_{g}cos\theta_{i}+cos\theta_{t}}\right)^{2}, \label{eqn9}
\end{eqnarray}
where $\theta_{i}$ is the incidence angle (with respect to the normal of PMT's glass), $n_{g}=1.5$ is PMT's glass refractive index and $sin\theta_{i}= n_{g}sin\theta_{t}$. These equations may be obtained from the Fresnel formulae (e.g.\ \cite{bor70}, p.\ 40, equations (21)).

\begin{figure*}[bt]
\begin{minipage}{0.47\textwidth}
\centering
\includegraphics[width=18pc]{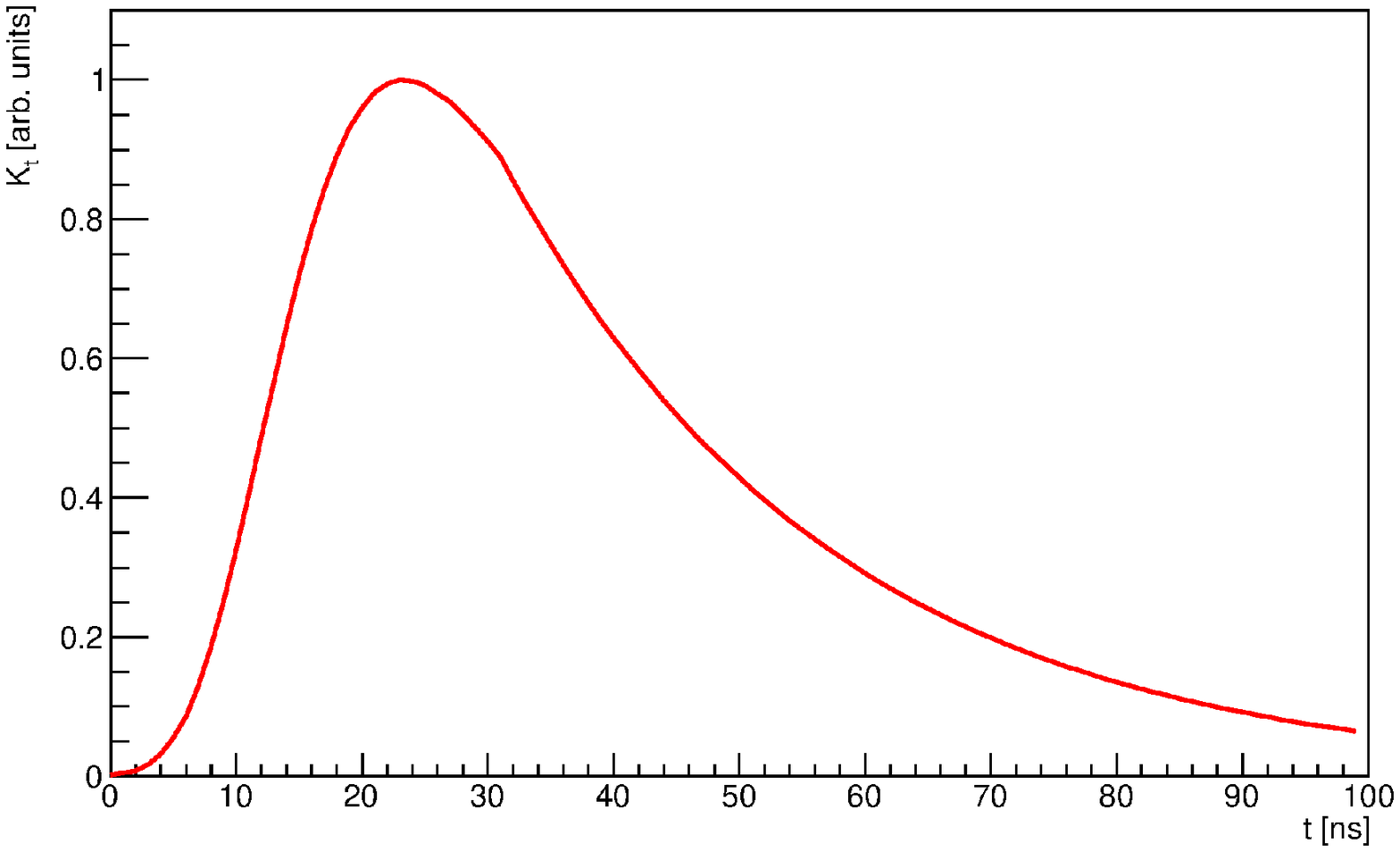}\hspace{2pc}%
\caption{The shape of time response function used in simulations (here it is normalized to the maximal value).}
\label{fig11}
\end{minipage}
\hfill
\begin{minipage}{0.47\textwidth}
\includegraphics[width=18pc]{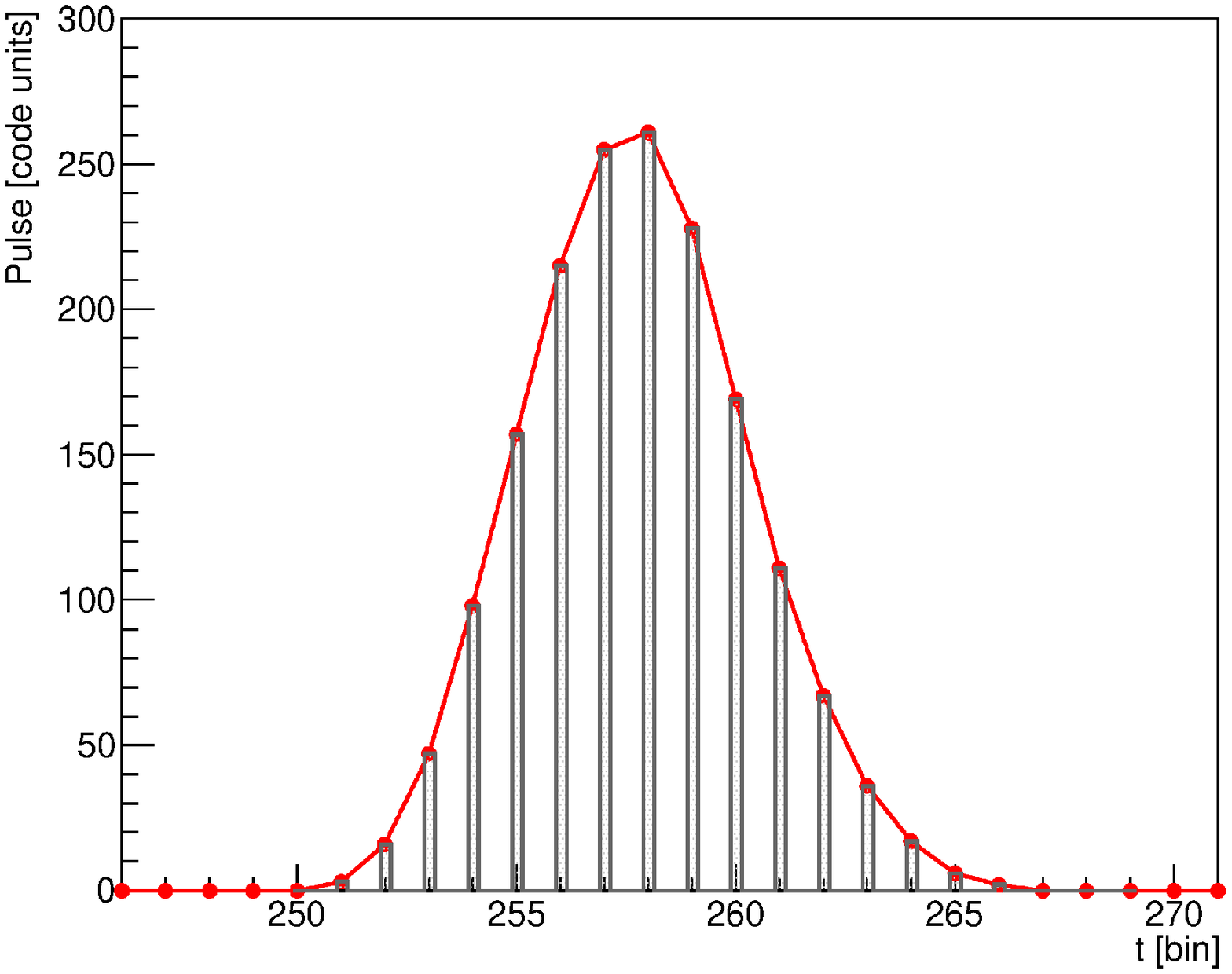}\hspace{2pc}%
\caption{Sketch of the ADC digitization procedure. Red line denotes $S_{out}(t)$ --- the shape of pulse after convolution with the PMT instrumental time response function, grey lines --- time intervals over which the ADC integration takes place, red circles --- typical measured amplitude $S_{D}$. One time bin equals to 12.5~ns.}
\label{fig12}
\end{minipage}
\end{figure*}
\suppressfloats[t]
\suppressfloats[b]

\begin{figure}[tb]
\centering
\includegraphics[width=13pc]{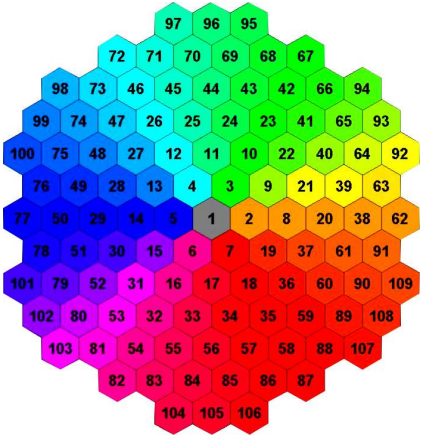}\hspace{2pc}%
\begin{minipage}[t]{16pc}\caption{The SPHERE-2 detector mosaic numbering scheme.}
\label{fig13}
\end{minipage}
\end{figure}

Diffraction of light was neglected. Polarization of light arriving at the diaphragm was set random. The direction of photons was determined by the coordinates of the detector and the scattering element of the snow surface (see above). Additional time delay which photons acquire when traced through the detector's optical scheme was also accounted for in the simulations, but it appears to be negligible. The intensity and time structure of noise is not known \textit{a priori}; therefore, as a first step, we have performed simulations of detector response without account of any noise. As well, here we do not take into account any additional broadening of pulse caused by the smearing effects of PMTs and electronics. These effects are accounted for in Subsect.~\ref{ssect:response-electronics}. For every photon hitting any PMT's photocathode the following quantities were recorded: \\
1. the number of this PMT \\
2. arrival time $t$ \\
3. impact distance, counted from the PMT's axis \\
4. incidence angle with respect to the PMT's axis. \\
Every simulated event represents an array of these 4 quantities.

To illustrate this stage of simulation we have estimated the discrete analogs of LDFs for \textit{Shower 1} and \textit{Shower 2} by direct counting of photons in every channel. The quantum efficiency of PMTs, it will be remembered, was already accounted for, thus the result of this procedure is already the array of photoelectron number emitted from PMT photocathodes. For practical purposes it is useful to associate the LDF value in every channel with some point on the snow surface. Here we assume that the coordinates of this point are such that photons originating from this point would hit the center of the photocathode of the corresponding PMT. These coordinates are shown in Fig.~\ref{fig9} for every channel by red circles for the case of $H$= 400 m. The discrete LDFs for \textit{Shower 1} and \textit{Shower 2}, again for the case of $H$= 400 m, are shown in Fig.~\ref{fig10} as black circles and green circles, respectively. Here we added a pedestal equal to 1 ph.el. to make the zero values visible.

\subsection{Response of the detector's electronics \label{ssect:response-electronics}}

The next stage of our simulation is the account of additional distortions of signal caused by the detector's electronics. The output of the \mbox{SPHERE-2} detector is measured in code units; therefore, in order to complete the simulation of detector response, we need to convert the array of photoelectrons to code units.

As the first step of this conversion procedure, for every measurement channel the model photoelectrons obtained at the previous step of simulation were histogrammed with time step 1.0~ns. After that, the broadening of pulse was accounted for by introducing the temporal convolution of simulated pulse shape with the PMT instrumental time response function:

\begin{equation}
S_{out}(t)= \int\limits_{0}^{\tau_{max}}{K_{t}(\tau)S_{inp}(t-\tau)d\tau} \label{eqn10},
\end{equation}
where $(S_{inp}(t), S_{out}(t))$ is the pulse before and after convolution, respectively, $\tau_{max}$= 100~ns and $K_{t}$ is the kernel (time response function) shown in Fig.~\ref{fig11}. The physical reason for such a broadening is the non-instantaneous time response of PMTs and electronics.

At the second step, simulation of the ADC digitization procedure was performed. The hardware used perform digitization every $dt_{ADC}= 12.5$ ns. Additionally, the process of digitization is not instantaneous, but takes place over a period of 3~ns. The result of the digitization procedure is a new array of values with time difference $dt_{ADC}$ between them, digitized values are calculated as $S_{D}= (S1+S2+S3)/3$, where $S1$, $S2$, and $S3$ are consecutive values of signal in initial model pulse with 1~ns step between them (see Fig.~\ref{fig12}).

\begin{figure*}[tb]
\begin{minipage}[h]{0.47\textwidth}
\centering
\includegraphics[width=18pc]{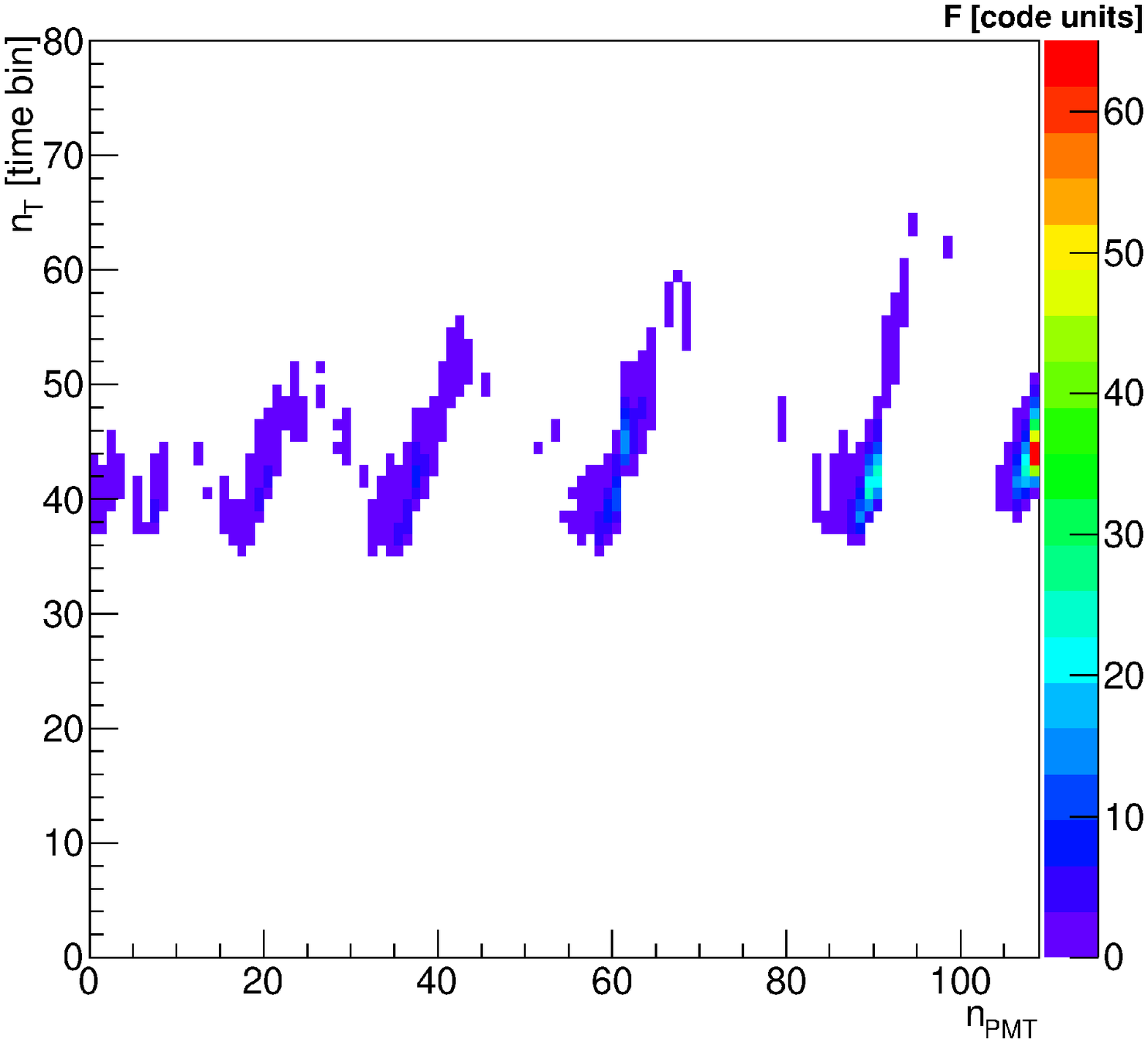}
\caption{Digitized response for \textit{Shower 1} without account of noise. PMT numbers $n_{PMT}$ follow the numbering scheme~\ref{fig13}.}
\label{fig14} 
\end{minipage}
\hfill
\begin{minipage}[h]{0.47\textwidth}
\centering
\includegraphics[width=18pc]{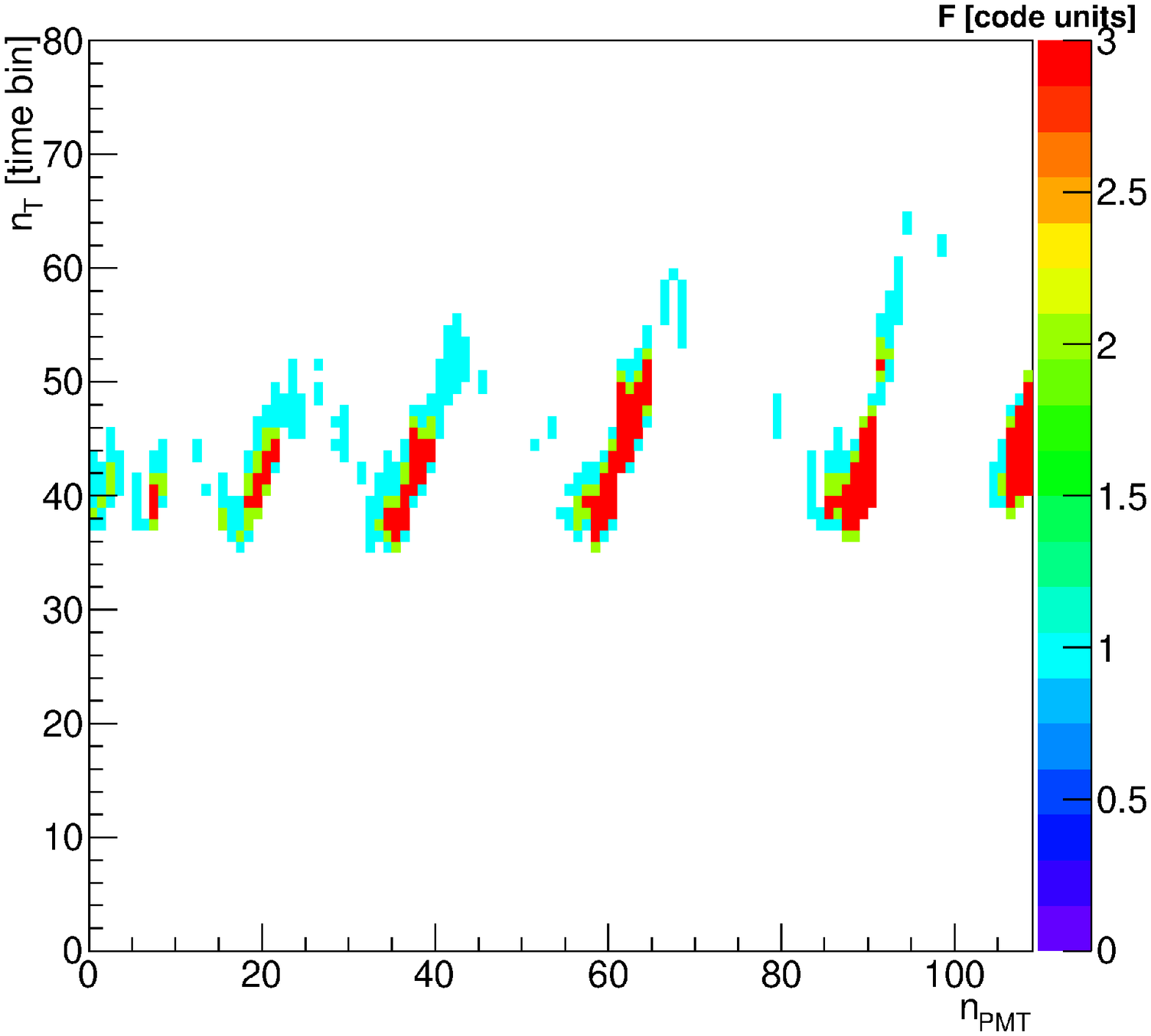}
\caption{The same, as in Fig.~\ref{fig14}, but all pixels with signal $S>3$ code units are set red.}
\label{fig15}
\end{minipage}
\end{figure*}

\begin{figure*}[tb]
\begin{minipage}[h]{0.47\textwidth}
\centering
\includegraphics[width=18pc]{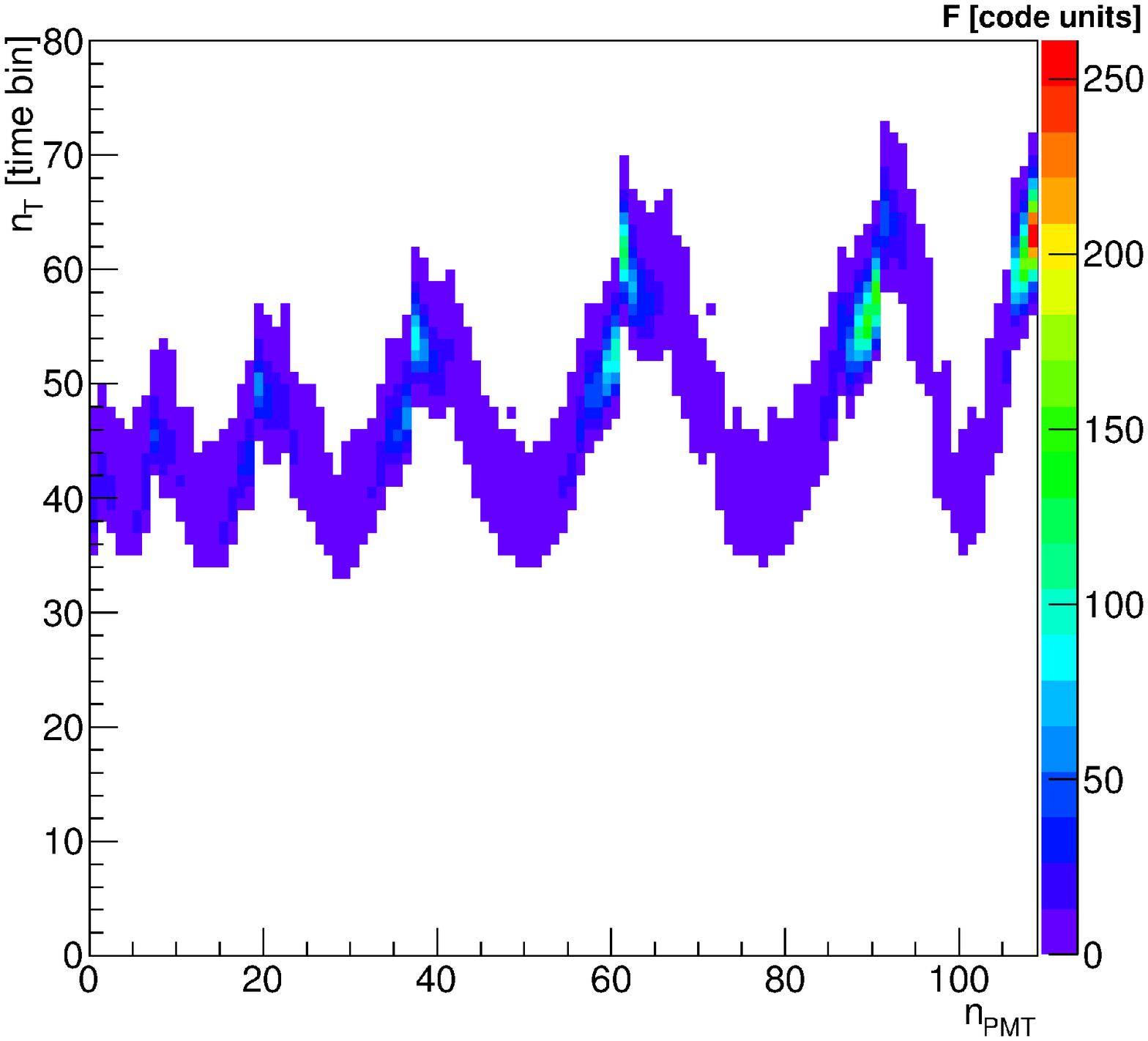}
\end{minipage}
\hfill
\begin{minipage}[h]{0.47\textwidth}
\centering
\includegraphics[width=18pc]{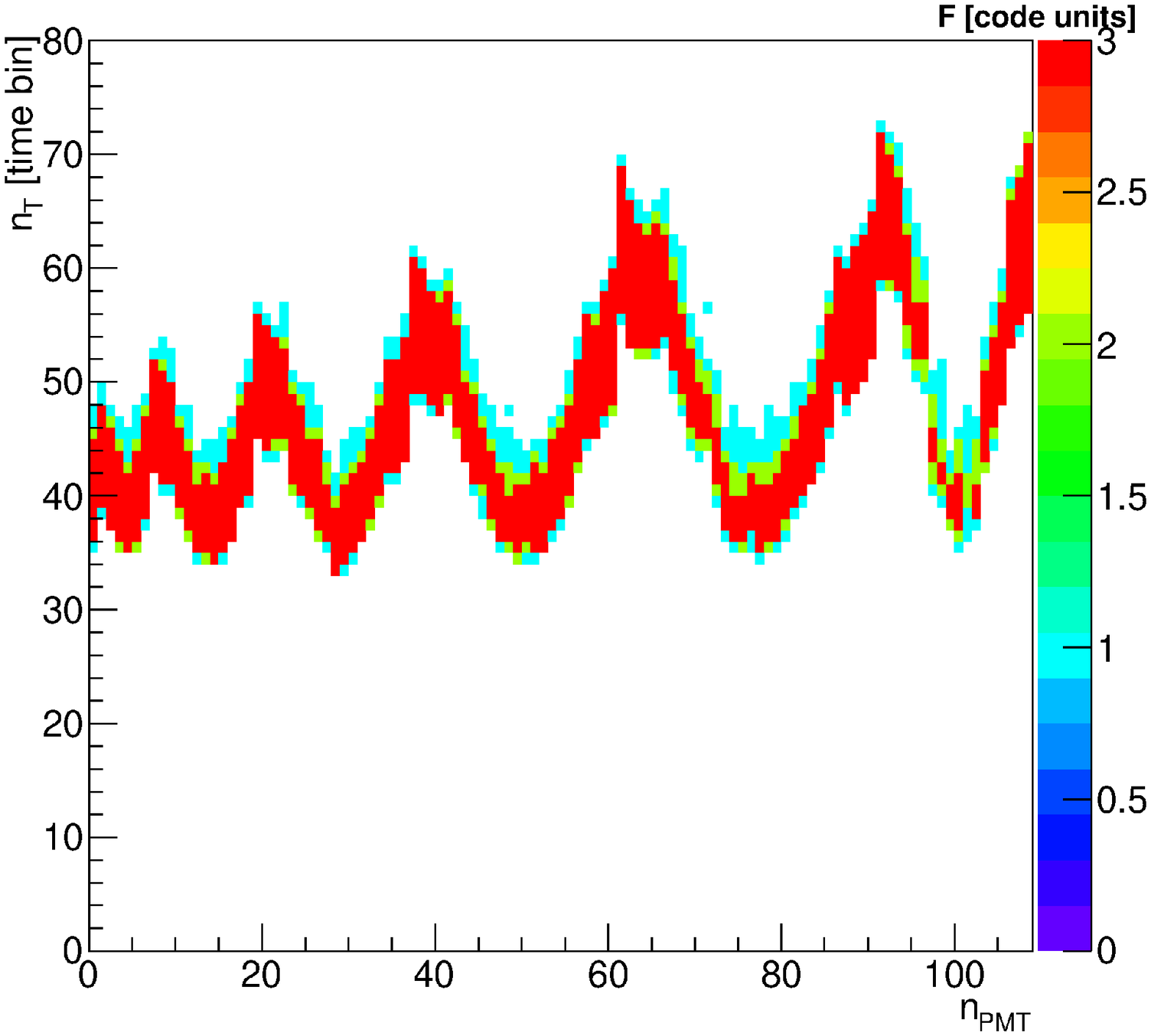}
\end{minipage}
\vfill
\begin{minipage}[h]{0.47\textwidth}
\centering
\caption{The same, as in Fig.~\ref{fig14}, but for \textit{Shower 2}.}
\label{fig16}
\end{minipage}
\hfill
\begin{minipage}[h]{0.47\textwidth}
\centering
\caption{The same, as in Fig.~\ref{fig15}, but for \textit{Shower 2}.}
\label{fig17}
\end{minipage}
\end{figure*}

At the third step, each bin of the latter array was multiplied to the factor $F_{abs}$ --- the photoelectron-to-code unit conversion factor. We set $F_{abs}$= 1.0, for this value is of the same order of magnitude as indicated by~\cite{ant16}. Finally, at the last step response arrays were rounded off to the nearest integer numbers. As a result of the described procedure, a two-di\-men\-sio\-nal array of response vs. channel number and time bin number was obtained for every model event.

The numbering scheme of the PMTs and measurement channels of the SPHERE-2 detector is shown in Fig.~\ref{fig13}. Several examples of digitized detector response are shown in Figs.~\ref{fig14}--\ref{fig17} for observation altitude $H$=400 m. Fig.~\ref{fig14} shows model response for \textit{Shower 1}. Response array's  bins content in code units is shown by color. The same response is shown in Fig.~\ref{fig15}, but for better visibility all bins with signal $S>3$ code units are set red. Similar two graphs for \textit{Shower 2} are shown in Fig.~\ref{fig16} and Fig.\ref{fig17} respectively. For response arrays presented in Fig.~\ref{fig14} and Fig.~\ref{fig16}, LDFs were calculated by direct summation of digitized signal over time bins. These LDFs are shown in Fig.~\ref{fig10} for \textit{Shower 1} and \textit{Shower 2} as red triangles and blue triangles respectively.

At high enough values of signal ($S>10$), LDF values calculated from photoelectron array (i.e. at the previous step of the simulation) and from digitized response are typically nearly identical. However, at low level of signal, $S<10$, both values are usually significantly different due to digitization effects. Indeed, the cells with the value of signal $S<0.5$ {code units} before digitization were rounded off to 0, thus causing the decrease of the computed LDF values. Sometimes, for long pulses, the digitization effect may appear significant even for LDF values exceeding 10 code units (compare the position of the last green circle and the last blue triangle at $R>$ 400~m in Fig.~\ref{fig10}). On the other hand, for the case of channels with sufficiently high signal values the digitization effect is small.

\section{Simulation of trigger response and instrumental acceptance \label{sect:acceptance}}

\begin{figure*}[p]
\begin{minipage}[h]{0.47\textwidth}
\centering
\includegraphics[width=16pc]{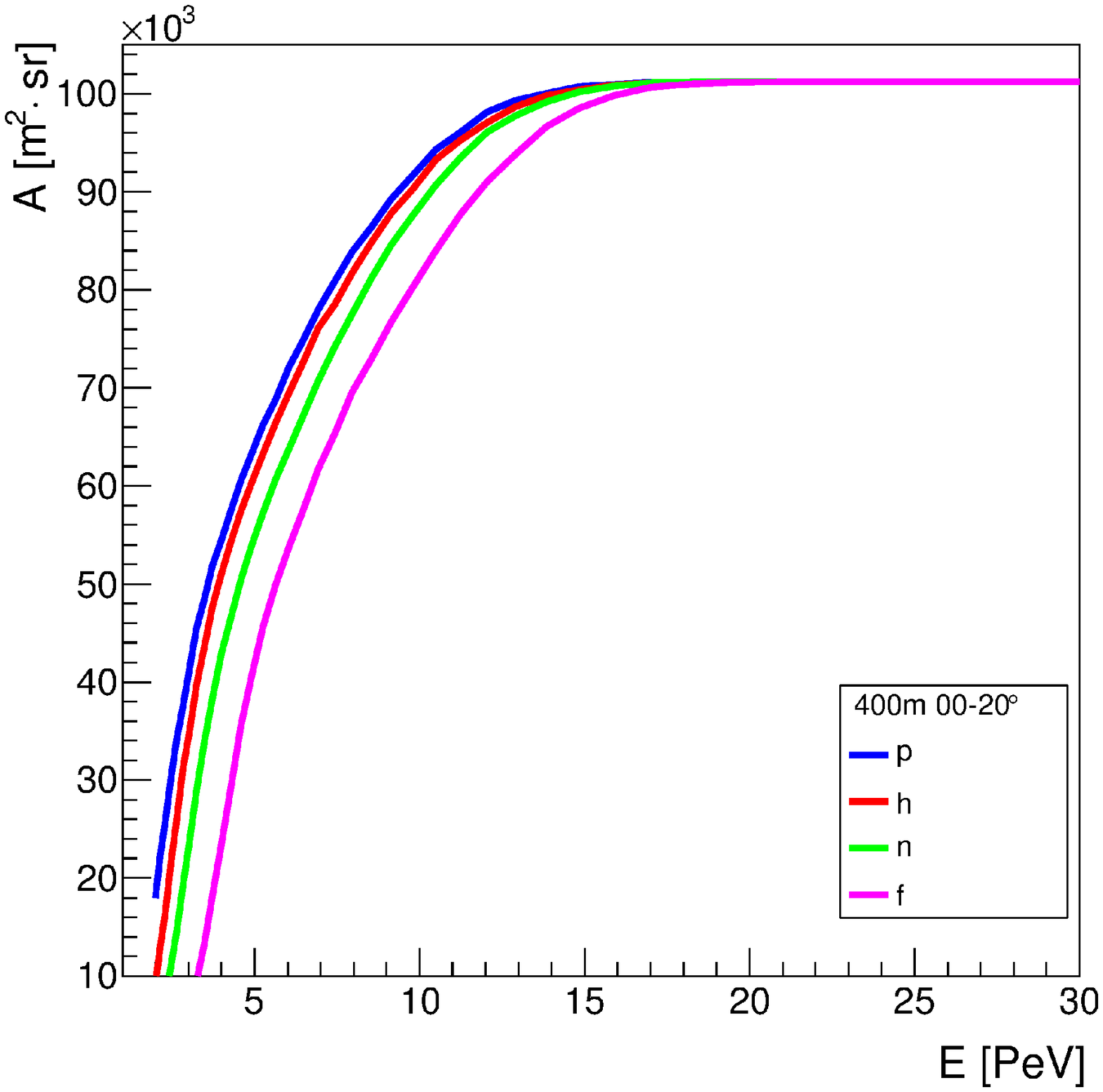}
\caption{Acceptance for various primaries, the angle range 0-20$^{\circ}$, and observation altitude 400~m. Primary protons: blue curve, primary Helium: red curve, primary Nitrogen: green curve, primary Iron: magenta curve.}
\label{fig18} 
\end{minipage}
\hfill
\begin{minipage}[h]{0.47\textwidth}
\centering
\includegraphics[width=16pc]{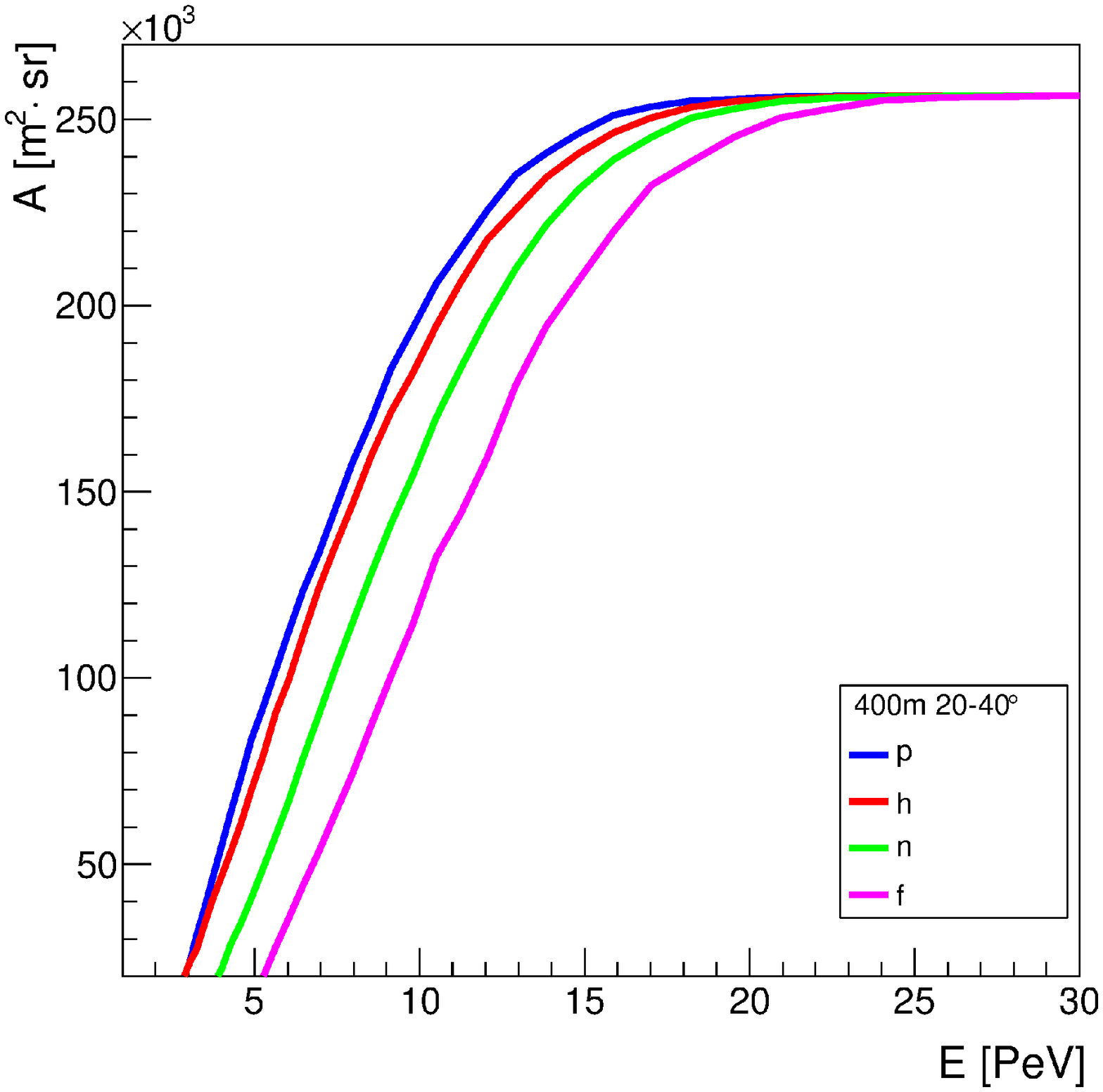}
\caption{The same, as in Fig.~\ref{fig18}, but for the angle range 20-40$^{\circ}$. Colors are the same as in Fig.~\ref{fig18}.}
\label{fig19}
\end{minipage}

\vfill

\begin{minipage}[h]{0.47\textwidth}
\centering
\includegraphics[width=16pc]{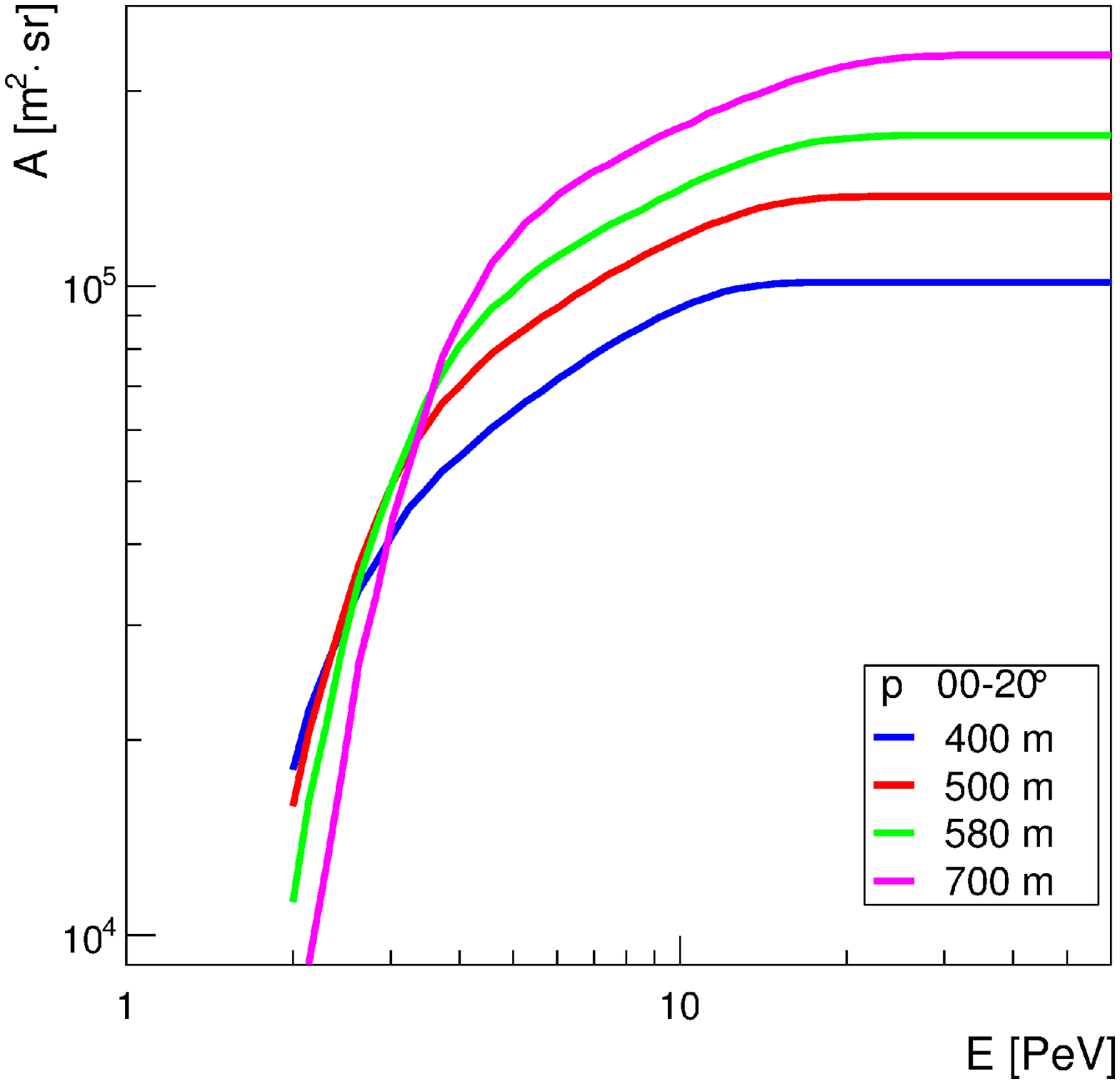}
\end{minipage}
\hfill
\begin{minipage}[h]{0.47\textwidth}
\centering
\includegraphics[width=16pc]{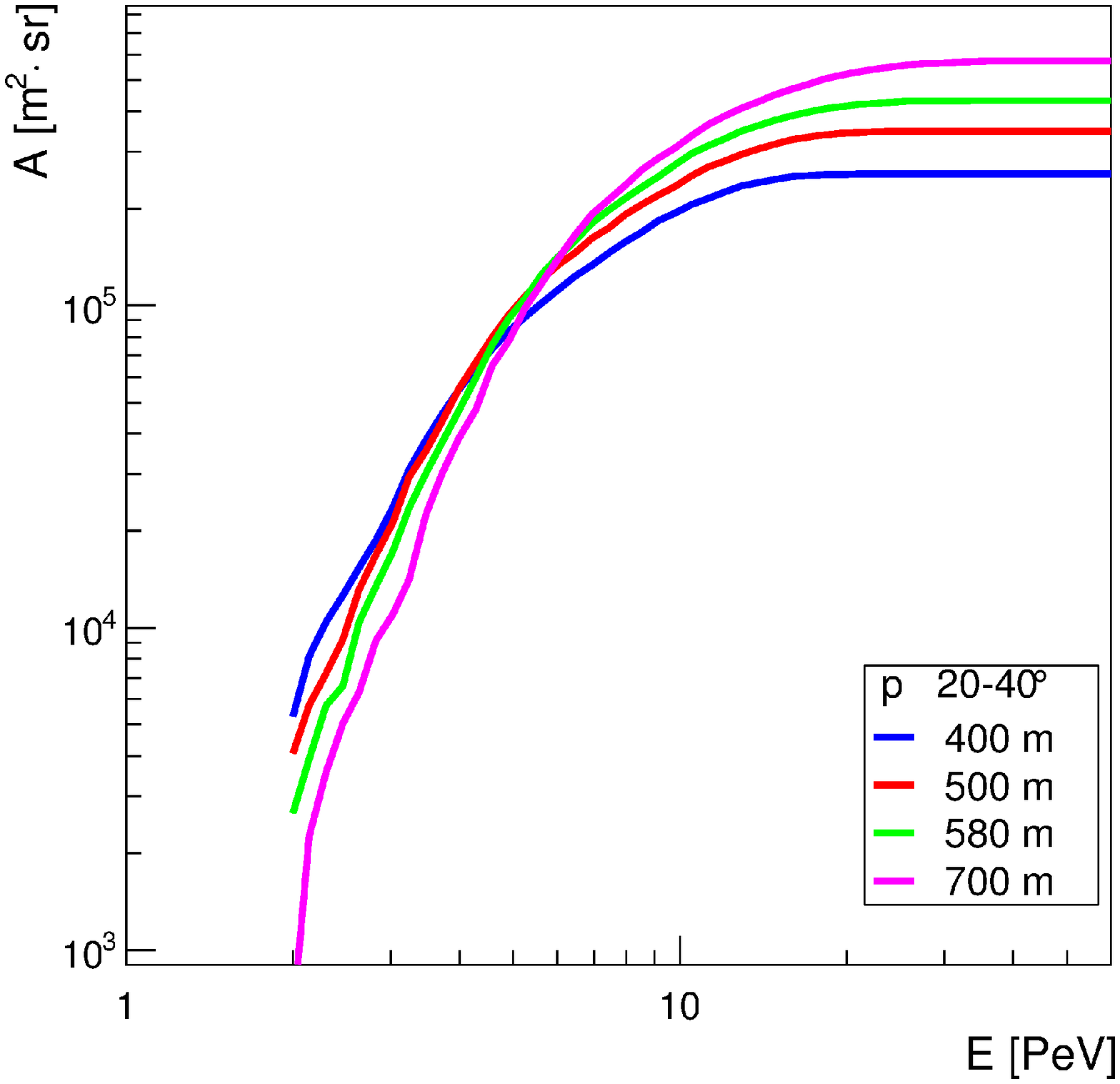}
\end{minipage}
\vfill
\begin{minipage}[h]{0.47\textwidth}
\centering
\caption{Acceptance for proton primaries and the angle range 0-20$^{\circ}$ for various altitudes: $H$= 400~m: blue curve, $H$= 500~m: red curve, $H$= 580~m: green curve, $H$= 700~m: magenta curve.}
\label{fig20}
\end{minipage}
\hfill
\begin{minipage}[h]{0.47\textwidth}
\centering
\caption{The same, as in Fig.~\ref{fig20}, but for the angle range 20-40$^{\circ}$. Colors are the same as in Fig.~\ref{fig20}.}
\label{fig21}
\end{minipage}

\vfill

\begin{minipage}[h]{0.47\textwidth}
\centering
\includegraphics[width=16pc]{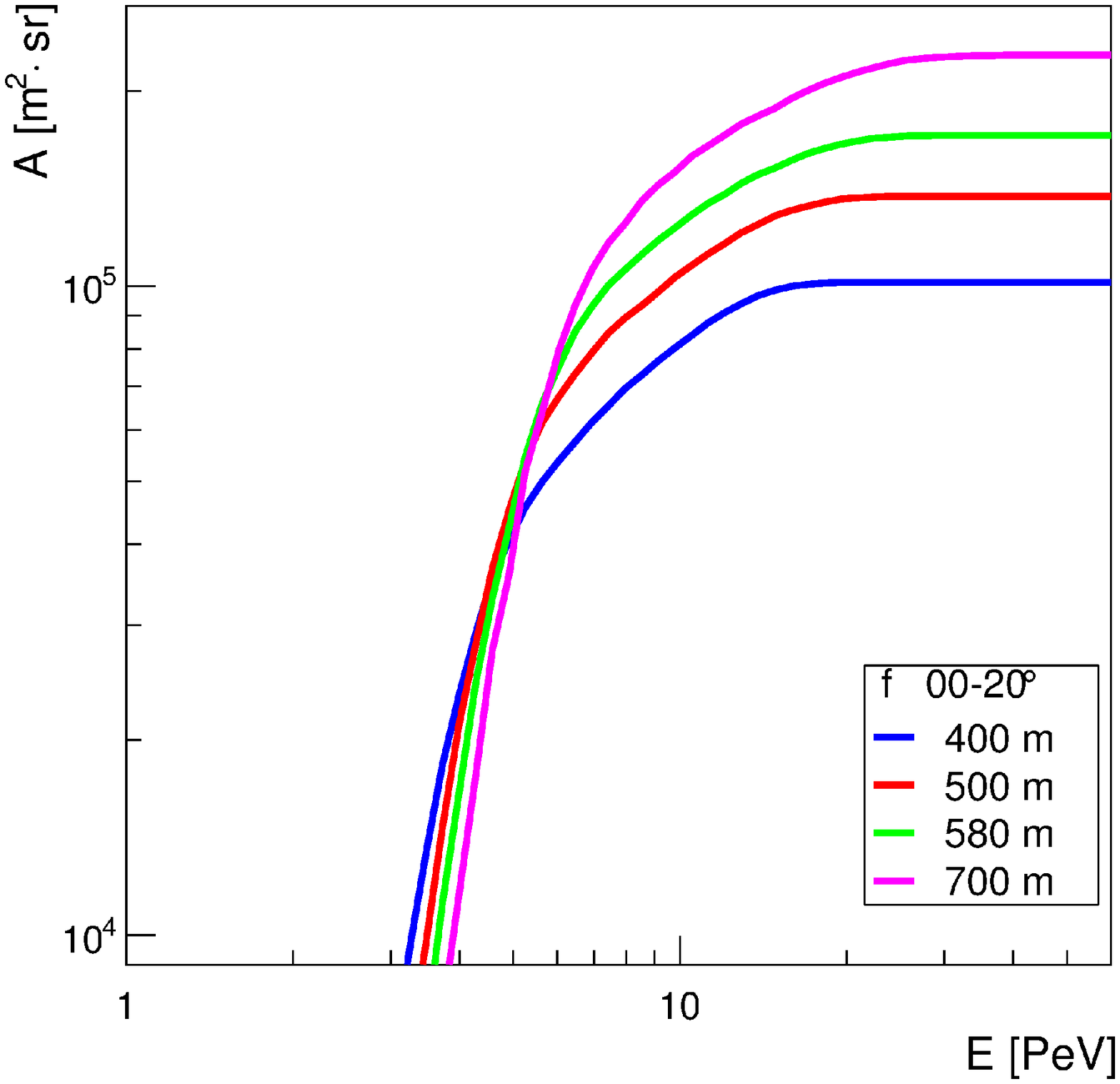}
\end{minipage}
\hfill
\begin{minipage}[h]{0.47\textwidth}
\centering
\includegraphics[width=16pc]{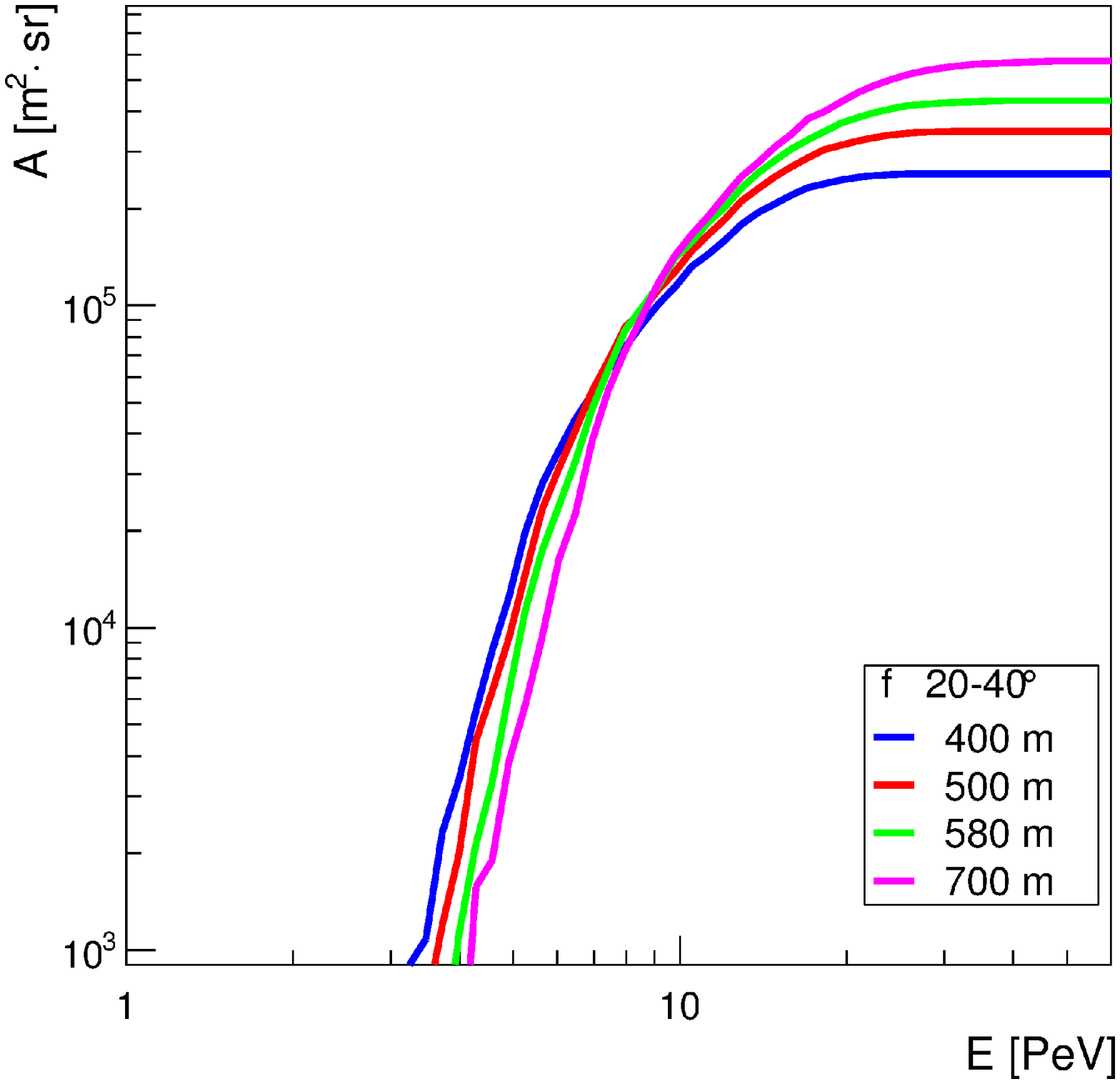}
\end{minipage}
\vfill
\begin{minipage}[h]{0.47\textwidth}
\centering
\caption{The same, as in Fig.~\ref{fig20}, but for Iron primaries.}
\label{fig22}
\end{minipage}
\hfill
\begin{minipage}[h]{0.47\textwidth}
\centering
\caption{The same, as in Fig.~\ref{fig21}, but for Iron primaries.}
\label{fig23}
\end{minipage}
\end{figure*}

Only a part of showers with emission observable by the \mbox{SPHERE-2} detector is actually recorded due to the restrictions imposed by the trigger conditions. These conditions were described in Sect.~\ref{sect:detector}. We performed a simulation of the trigger response for a large sample of model showers. This sample included four various primaries: proton, Helium, Nitrogen, and Iron with zenith angles in the two separate ranges of 0--20$^{\circ}$ and 20--40$^{\circ}$, for several altitudes $H=(400, 500, 580, 700)$~m. The axis coordinates of these showers were distributed uniformly inside a square with dimensions 1.5 $H$ $\times$ 1.5 $H$. Every event taken from the CORSIKA simulation was repeatedly used to simulate detector response, as described in Sect.~\ref{sect:response}, 100 times with different axis coordinates filling the above-described square.

Besides the detector response, the response of the trigger for every model shower depends also on the detector amplitude threshold values, calibration factors, and noise in every channel. The thresholds, it will be remembered, are set at the beginning of each flight (see Sect.~\ref{sect:detector}), and, if the data taking was interrupted for some reason, also at the beginning of every new start of data taking. Both threshold and noise arrays are recorded by the data acquisition system and include the same instrumental pedestals.

It is desirable that the trigger bit value $B_{T}$ is calculated for a large set of primary energies; however, by now our simulations were performed for 10~PeV, 30~PeV, and 100~PeV only. Fortunately, the amplitude of EAS is almost proportional to the primary energy \cite{hil82,pat83,ded04}, therefore $B_{T}$ for the primary energy $K\cdot E$ may be estimated by applying the same procedure as for the case of the primary energy $E$, but substituting $K\cdot F(nx,ny,nt)$ instead of $F(nx,ny,nt)$ as the input function.

\begin{figure*}[t]
\begin{minipage}[h]{0.47\textwidth}
\includegraphics[width=19pc]{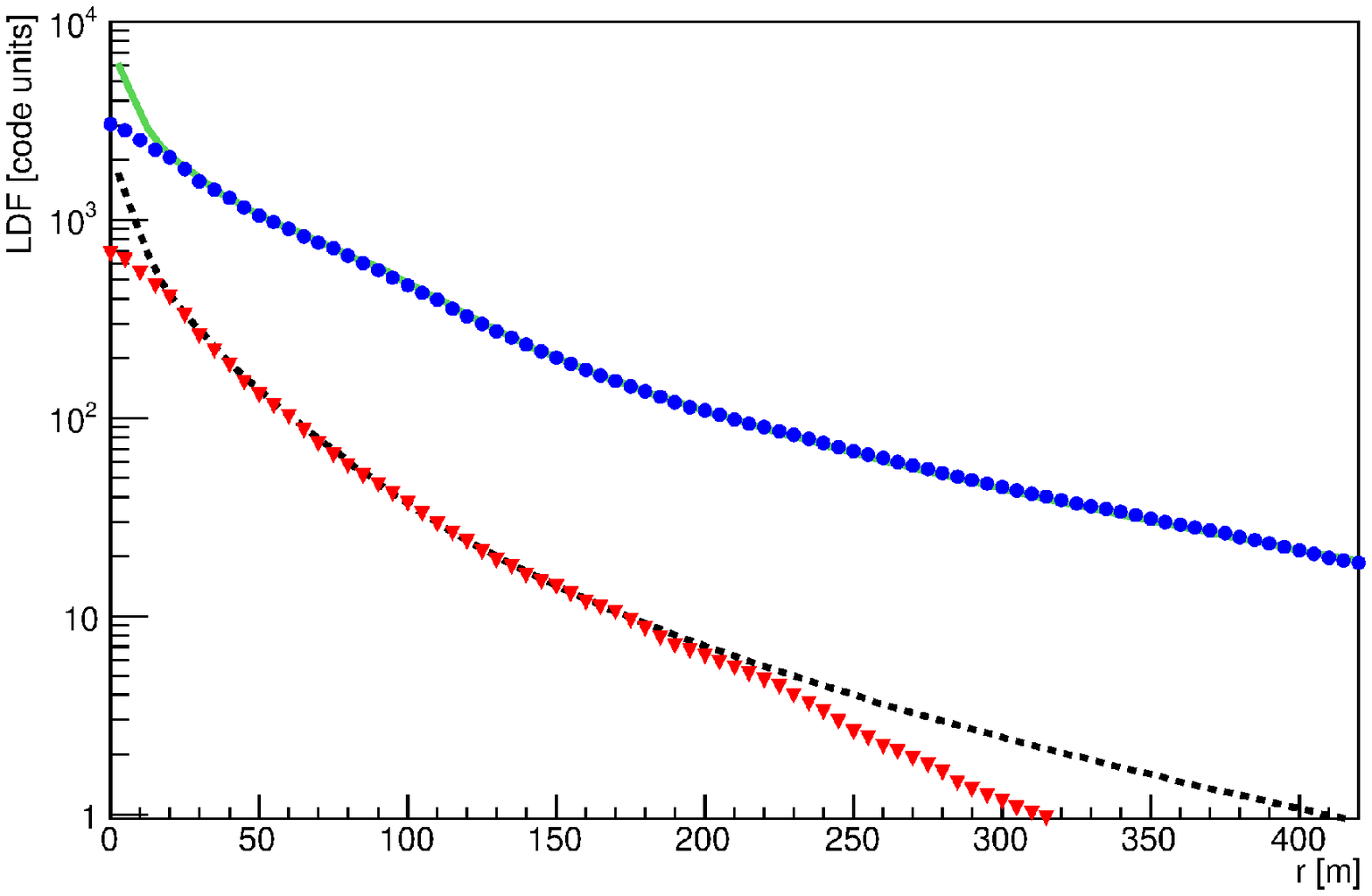}\hspace{1pc}%
\centering
\caption{Examples of composite model LDF for \textit{Shower 1} (red triangles) and \textit{Shower 2} (blue circles) superimposed on the corresponding model LDFs at the ground level (black dashed curve and green solid curve, respectively).}
\label{fig24}
\end{minipage}
\hfill
\begin{minipage}[h]{0.47\textwidth}
\centering
\includegraphics[width=20pc]{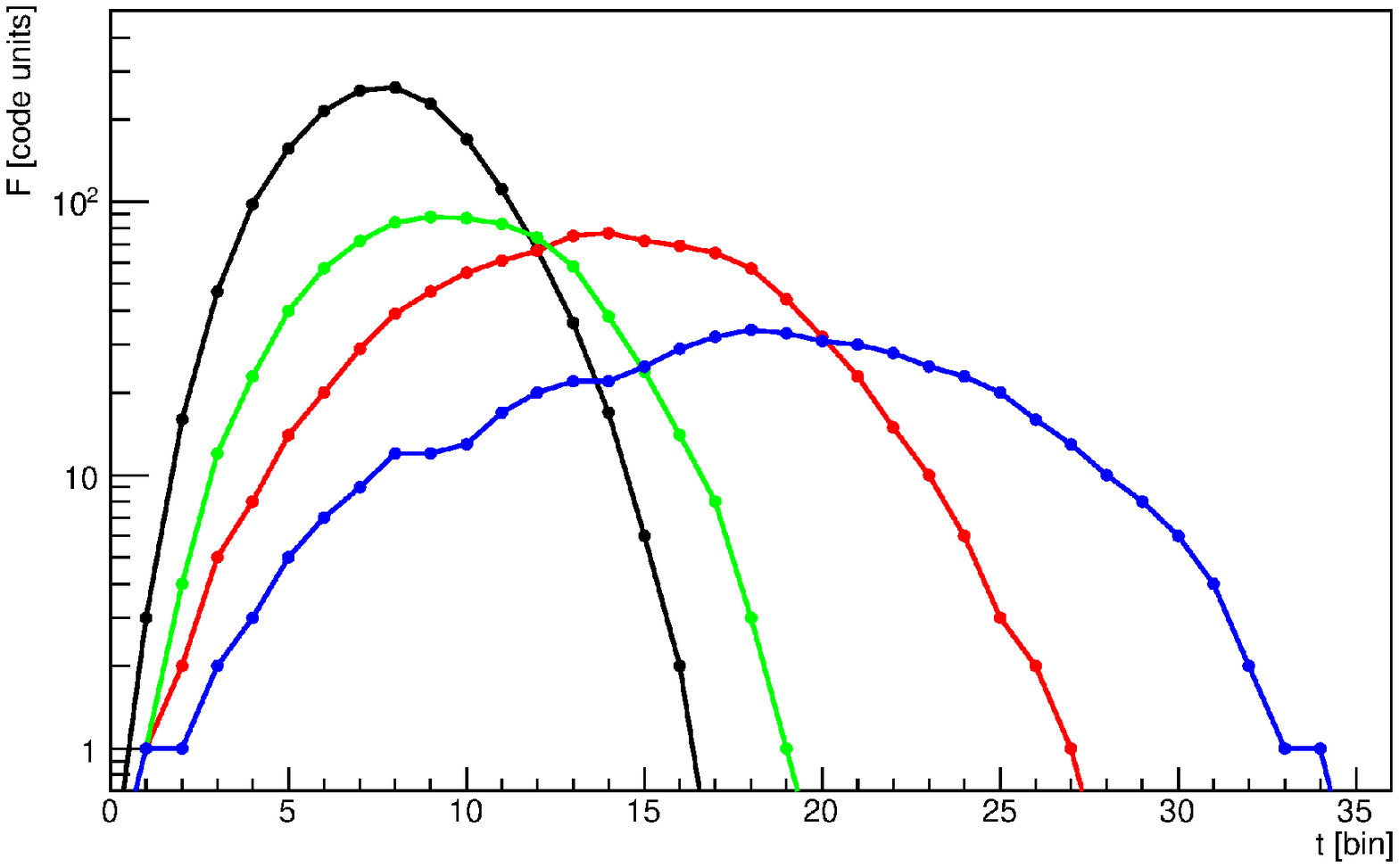}\hspace{1pc}%
\caption{Several examples of individual model pulses: \textit{Shower 2} and  $H$= 400~m (black), \textit{Shower 2} and  $H$= 900~m (red), \textit{Shower 3} and  $H$= 400~m (green), \textit{Shower 3} and  $H$= 900~m (blue). One time bin equals to 12.5~ns.}
\label{fig25}
\end{minipage}
\end{figure*}

The arrays of $B_{T}(E)$, calculated separately for each flight, were used as input functions to estimate the instrumental acceptance $A(E)$. The fiducial area for this procedure is a circle with the radius $R_{A}= R_{FOV}(H)+100$ m. The results of such calculations for various primaries, observation altitudes, and zenith angle ranges are presented in Figs.~\ref{fig18}--\ref{fig23}. Fig.~\ref{fig18} shows that for nearly-vertical EAS ($\theta<$20$^{\circ}$) the acceptance saturates below $E\approx$18~PeV irrespectively of the mass of the primary nucleus. The $A(E)$ dependence saturates faster for lighter nuclei; in this latter case the effective threshold also appears to be somewhat less than for heavier nuclei. For the case of larger zenith angles ($20^{\circ}<\theta< 40 ^{\circ}$) (Fig.~\ref{fig19}) the dependence of acceptance on energy is qualitatively the same, but $A(E)$ saturates at somewhat higher energy, $E\approx25$~PeV. These features for heavier primary nuclei and larger zenith angles are naturally explained by the fact that for these cases the signal amplitude near the axis, $R<40$-50~m, is smaller due to earlier development of showers in terms of vertical depth in the atmosphere. 

The dependence of acceptance on the observation altitude for primary protons and $\theta<20^{\circ}$ is shown in Fig.~\ref{fig20}. For higher altitudes, the saturation energy and the effective energy threshold are also higher. This is qualitatively explained by the fact that for higher altitude the trigger response depends on larger span of distances, thus requiring higher energy for triggering, because Cherenkov light LDF, as a rule, quickly falls with distance from the axis. The same dependence, but for $20^{\circ}<\theta<40^{\circ}$, is presented in Fig.~\ref{fig21}.  Fig.~\ref{fig21} shows qualitatively the same features as Fig.~\ref{fig20}, but, again, the overall picture is shifted towards higher energies. Finally, the same figures as Figs.~\ref{fig20}--\ref{fig21}, but for Iron primaries, are presented in Figs.~\ref{fig22}--\ref{fig23}.

We note that reconstruction inefficiencies would result in a decrease of the number of registered EAS. In order to estimate the magnitude of this effect, we performed reconstruction of model events with realistic background taken directly from experimental data. The typical inefficiency was found to be rather low, less than 2-3 \% at $E$=10 PeV and even smaller ($\sim$1 \%) for $E$=30 PeV and 100 PeV.

\section{Composite model quantities and the spatial and temporal structure of observable signal \label{sect:composite}}

Some characteristics of model showers were already shown in Figs.~\ref{fig10}, \ref{fig14}--\ref{fig17}, and briefly discussed. However, fluctuations of individual shower signal are considerable, even for the case of 100 PeV showers (see Fig.~\ref{fig10}). Therefore, to understand distortions caused by the \mbox{SPHERE-2} detector, it would be useful to introduce some quantity that represents an average signal for a certain EAS with account of detector's response, but with reduced statistical fluctuations.

To meet these requirements, we devised and put to use the following procedure. We already have calculated the sample of model showers needed for trigger response simulation. Now we perform the averaging over every 100 response events (after the digitization stage) that originate from the same EAS simulated with the CORSIKA code to compute average, ``composite'' quantities, such as the ``composite model LDF'' (CLDF) and ``composite model pulse'' (CMP) to make subtler features of model showers more readily identifiable.

While calculating composite model LDFs, the ``cylindrical projection'' procedure is applied for every shower, similarly to the procedure described in Sect.~\ref{sect:ground}. As well, some additional corrections are introduced as follows. As indicated in eq.~\ref{eqn3}, the number of photons, radiated from the element of snow surface, depends on the coordinates $(x,y)$ and $\theta_{n}$ as $cos^{2} \theta_{n}/(H^{2}+dr^{2})$, where $dr= \sqrt{x^{2}+y^{2}}$. On the other hand, the area of the snow surface that corresponds to the PMT field of view (FOV) relative to the same quantity for the central PMT  is roughly proportional to $H^{2}+dr^{2}$. Therefore, the factor $H^{2}+dr^{2}$ cancels out, and the final LDF compensation factor equals to $cos^{2}\theta_{n}$. We introduce this last correction for every individual model LDF, and then perform the averaging over 100 realisations of individual model LDFs in spatial bins 5~m wide (counted from the axis of every shower) to obtain the corresponding composite model LDF. Finally, we perform some smearing over distance to the shower's axis using a kernel with width growing vs. this distance.

Likewise, for every shower taken from the CORSIKA simulation a set of composite model pulses was calculated. Here we are interested in dependence of the CMP shape on the position of measurement channel in the mosaic, as well as the distance from the shower's axis. Therefore, the CMP set for every CORSIKA shower is represented by a four-dimensional array $P[nring][n\phi][nr][nt]$ with bins on $nring$ --- the number of the ring of a PMT in mosaic counted from the center of the mosaic, $n\phi$ --- azimuthal angle of the PMT, $nr$ --- the bin number on distance from the shower's axis to the PMT, and $nt$ --- the time bin number.

\begin{figure*}[bt]
\begin{minipage}[h]{0.47\textwidth}
\centering
\includegraphics[width=18pc]{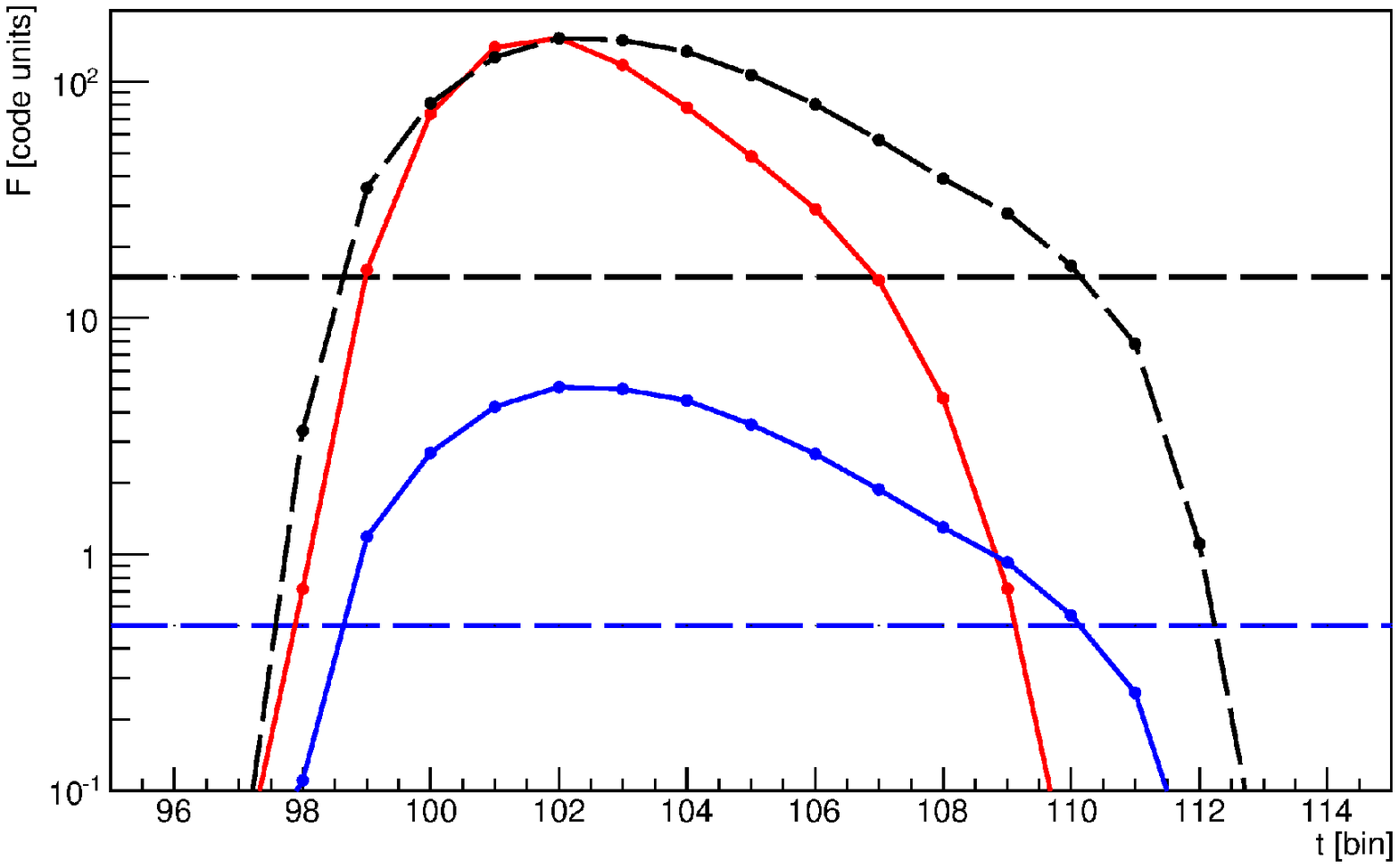}
\caption{The dependence of composite model pulse shape on the distance from the center of the detector's FOV to the center of the projection of the PMT's FOV to the snow surface for \textit{Shower 2}. Two composite model pulses in channels 0 (red) and 108 (blue) are shown. Black circles together with enveloping dashed black lines represent blue curve normalized to the maximum of red curve. One time bin equals to 12.5~ns.}
\label{fig26}
\end{minipage}
\hfill
\begin{minipage}[h]{0.47\textwidth}
\centering
\includegraphics[width=18pc]{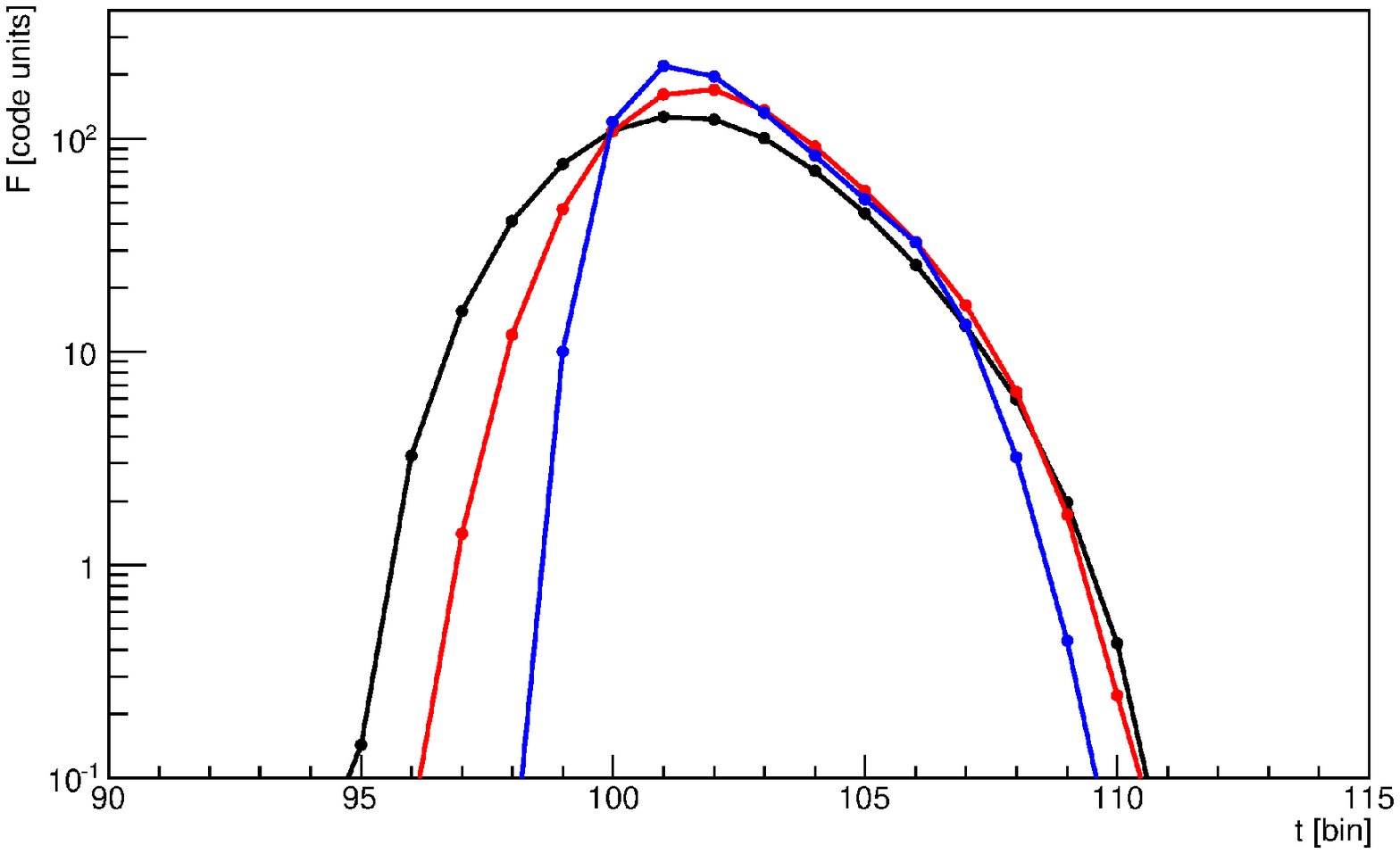}
\caption{Dependence of composite model pulses on the $\phi_{r}$ angle: $\phi_{r}$=0 (black), $\phi_{r}$= $\pi/2$ (red), $\phi_{r}$= $\pi$ (blue). One time bin equals to 12.5~ns.}
\label{fig27}
\end{minipage}
\end{figure*}

Two examples of calculated composite model LDF for \textit{Shower 1} and \textit{Shower 2} are presented in Fig.~\ref{fig24} together with initial LDF before the detector response simulation stage. The fluctuations of these composite model LDFs are greatly suppressed with respect to the corresponding individual model LDFs (compare with Fig.~\ref{fig10}). Model LDFs at the ground level (``initial LDFs'') shown in Fig.~\ref{fig24} were taken from the CORSIKA simulation, subjected to the same cylindrical projection procedure, and normalized to the corresponding composite model LDFs. For the case of primary energy 10~PeV, the initial and composite LDF practically coincide for the distance range $R$=20--180~m. The composite LDF is below the initial one for $R<20$~m due to a blurring effect introduced by the optical system of the \mbox{SPHERE-2} detector. At $R>180$~m another important effect causes the drop of the composite LDF below the initial one, namely, the digitization effect already discussed in Subsect.~\ref{ssect:response-electronics}. For the case of \textit{Shower 2} the blurring effect is, similarly, at place, because it does not depend on the amplitude of the signal, but the digitization effect is absent at $R<400$~m due to higher primary energy of \textit{Shower 2}, and, consequently, higher amplitude at the same value of $R$ compared to the case of \textit{Shower 1}. Once again, for the case of moderate zenith angles of \textit{Shower 1} and \textit{Shower 2} the digitization effect is, as a rule, negligible if the amplitude of the signal $S$ exceeds 10 code units, in agreement with values indicated in Subsect.~\ref{ssect:response-electronics}.

Likewise, the properties of individual and composite model pulses were studied, see Figs.~\ref{fig25}--\ref{fig27}. Fig.~\ref{fig25} shows the dependence of individual model pulse shape in certain measurement channel for two detector response events vs. zenith angle and observation altitude. In this case both showers had the same axis position, and we chose to present the pulse shape in the channel number 108; in this particular case it appeared to be the nearest to the axis. The higher the zenith angle of the primary particle and the altitude, the longer the pulses are. It is important to note that the shape of the pulse is determined not only by the properties of the primary nucleus (mostly by the zenith angle), but also by the observation conditions, in particular, by the altitude.

Another important condition that influences the shape of the pulse is the position of the measurement channel in the mosaic, as illustrated in Fig.~\ref{fig26}. Equations~\ref{eqn4}--\ref{eqn6} define time delay of photons arriving to the detector that produce photoelectrons and determine the shape of observable pulse (measured in code units). The quantities $d(t_{3})/dx$ and $d(t_{3})/dy$ grow with $x$ and $y$, respectively, causing longer pulses in outer channels due to the fact that PMT field of view projections on the ground cover certain areas with considerable spatial extension. Another factor that to the lesser extent broadens the pulse is the growing extension of outer PMTs field of view. Fig.~\ref{fig26} also illustrates the digitization effect on the pulse that causes an artificial cutoff of observable pulses below the amplitude of 0.5 code units  (this level is denoted by horisontal dashed blue line). The same effect is even better visible in the re-normalized version of the same pulse (black circles and enveloping dashed black lines);  in this case the re-normalized value of the 0.5 code units level is shown by horizontal dashed black line.

Finally, Fig.~\ref{fig27} shows the dependence of composite pulse shape vs. $\phi_{r}$ defined as the azimuthal angle between the following two unit vectors $\vec{d_{0}}$ and $\vec{d}$ on the ground surface. Both these vectors have their origin at the shower axis position. The direction of $\vec{d_{0}}$ is defined by the projection of the primary particle direction to the ground level, the direction of $\vec{d}$ --- by the coordinates of the center of particular measurement channel FOV. This dependence is caused by the fact that the $t_{2}$ and $t_{3}$ terms may interfere constructively or destructively depending on the $\phi_{r}$ value (see equations~\ref{eqn4}--\ref{eqn6}). Thus, the pulse shape acquires a complex dependence on the $\phi_{r}$ value. To conclude, the shape of observable pulse depends on many parameters in a non-trivial way, once again highlighting the importance of detailed simulations for subsequent data analysis. In total, we have obtained a large sample of composite model LDFs (several thousand) and composite model pulses (several hundred thousand) in order to account for fluctuations of EAS development.

\begin{figure*}[t]
\begin{minipage}[h]{0.47\textwidth}
\includegraphics[width=19pc]{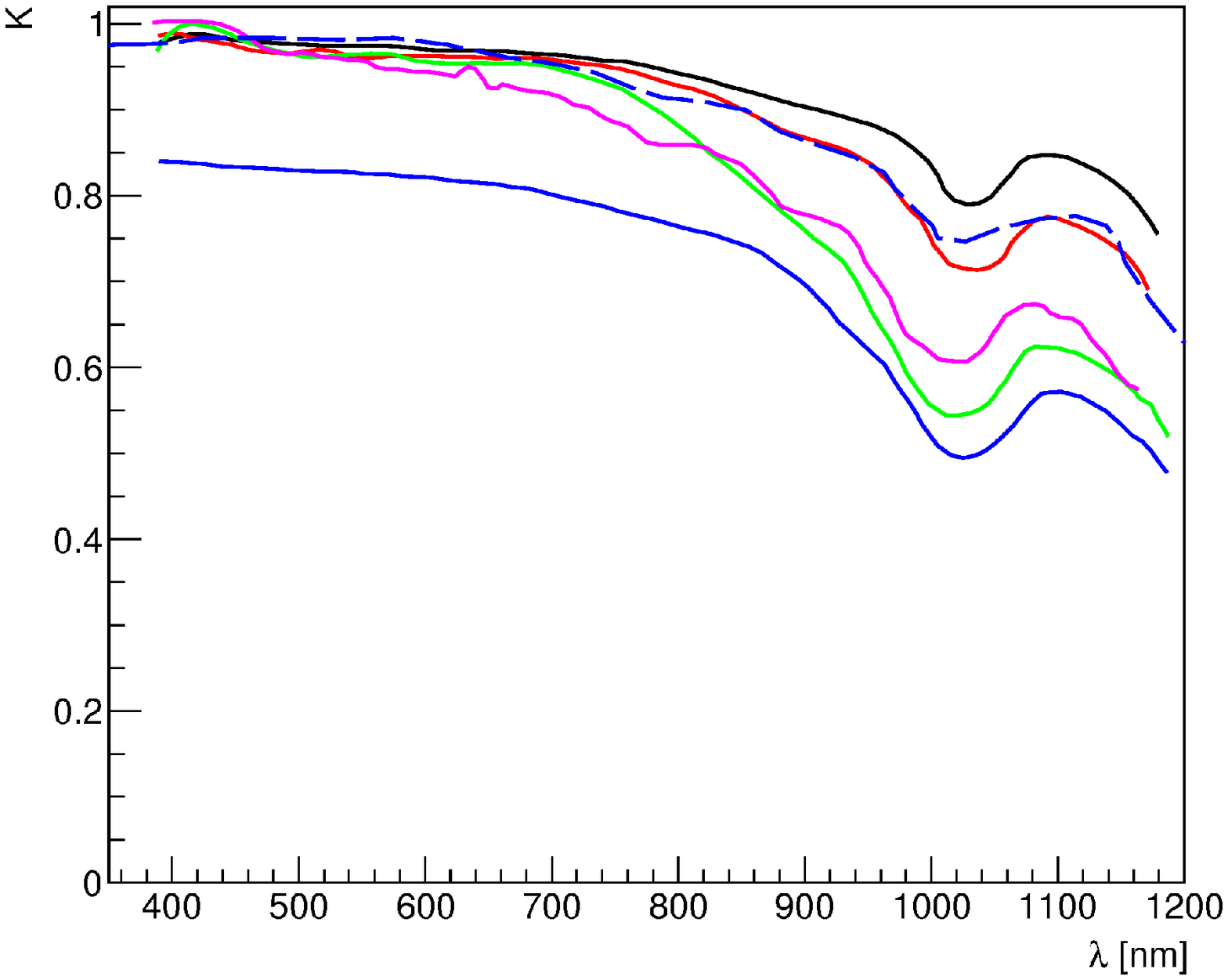}\hspace{1pc}%
\centering
\caption{Dependence of snow albedo on the wavelength according to various measurements. Solid curves show results from~\cite{qun83}, dashed blue curve --- from~\cite{gre94} (more details in the text).}
\label{fig28}
\end{minipage}
\hfill
\begin{minipage}[h]{0.47\textwidth}
\centering
\includegraphics[width=19pc]{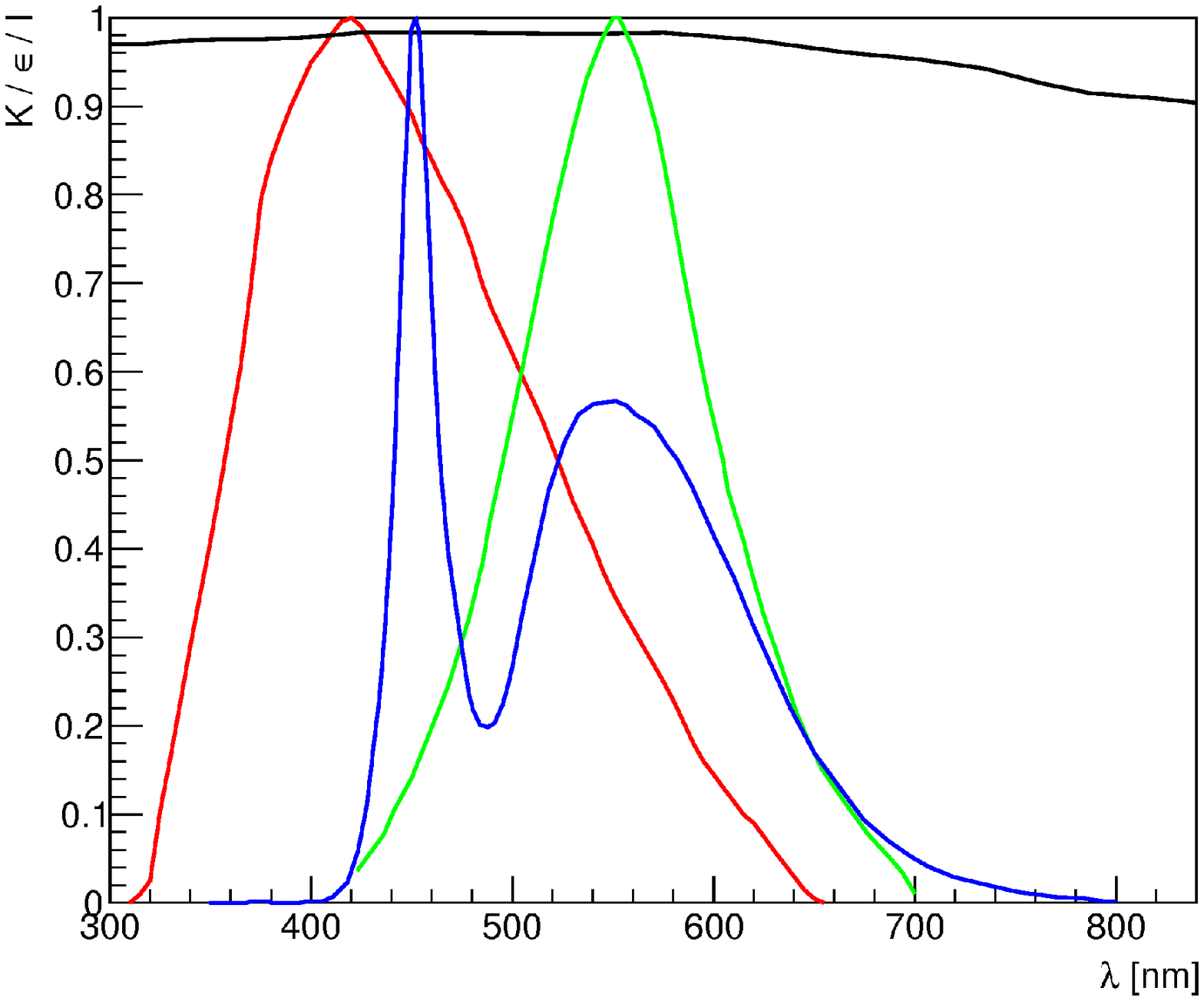}\hspace{1pc}%
\caption{Black curve denotes albedo from~\cite{gre94}. Other curves are normalized to unity at the maximum. Red curve denotes relative quantum efficiency of PMT-84-3, green curve --- relative sensitivity of the luxmeter, blue curve --- relative spectrum of the light source used for the BRDF measurements (for more details of these measurements see Subsect.~\ref{ssect:snow-brdf}).}
\label{fig29}
\end{minipage}
\end{figure*}

\section{The impact of realistic snow optical properties on model events \label{sect:snow}}

In Sect.~\ref{sect:response} we have assumed a simplified model of reflection from the snow cover, universally known as isotropic or Lambertian reflection \cite{bor70}. Here we review contemporary knowledge on snow optical properties and study the impact of these on simulated EAS detector response events.

\subsection{Snow albedo \label{ssect:snow-albedo}}

Several measurements of snow albedo $K$ vs. the wavelength $\lambda$ under various conditions are presented in Fig.~\ref{fig28}. Black, red, green and blue solid curves describe the time evolution of albedo of a particular snowpack according to~\cite{qun83}. These measurements were performed in the north of China, actually not far from our experimental site. Black solid curve denotes the albedo of fresh snow, which is typically the case for our observations. Red solid curve corresponds to wet snow. This is not typical for our observations, but such a situation still may occur sometimes. Green and blue curves denote snow with severe metamorphism; such a situation was never observed during our measurement runs at Lake Baikal. Solid magenta curve corresponds to the case of fresh snow and another sample of snow. In this case the measurements yielded the result that is very close to unity at $\lambda$= 400-450 nm. This may be explained by systematic uncertainties while compensating for geometrical effects (for instance, see~\cite{gre94}). Finally, dashed blue curve denotes the albedo for fresh Antarctic snow from~\cite{gre94}. Except for the case of severely metemorphosed snow (solid blue curve) all presented measurements show a very high value for albedo, $K>$0.9 for the case of $\lambda=$400-600 nm. Moreover, the difference between these results does not exceed several percent.

We have checked that the results of other available measurements of snow albedo, such as~\cite{dum10}, are close to those shown in Fig.~\ref{fig28}. For instance, the difference of the results on albedo of~\cite{hud06} and~\cite{gre94} (dashed blue curve in Fig.~\ref{fig28}) is less than 1 \% for $\lambda$=400--900 nm. The results of~\cite{dum10} show nearly the same shape of albedo vs. wavelength as \cite{gre94}, but for the case of wet snow and normal incidence $K\approx$0.9. These results are not shown in Fig.~\ref{fig28} to avoid the confusion of this graph with too many curves. To conclude, snow albedo in the 300-600 nm wavelength region is very high ($K\approx$0.90-0.98) and weakly dependent on snow physical conditions. In contrast, in the infra-red (IR) wavelength region ($\lambda>$800 nm) $K$ is usually somewhat lower than in the optical band and may be as low as 0.5-0.6 for wet snow at $\lambda\approx$1000 nm due to absorption.

Now let us demonstrate what wavelength region is of actual interest to our experiment. Fig.~\ref{fig29} shows a fragment of the $K(\lambda)$ dependence according to~\cite{gre94} (i.e. a sample from the same dataset as in Fig.~\ref{fig28}, dashed blue curve) together with the PMT-84-3 quantum efficiency $\epsilon(\lambda)$ normalized to unity at the maximum (the actual value at the maximum is 0.2). Obviously, the range of interest on wavelength for our work is 300-650 $nm$, where albedo is high and well known for fresh snow.

The temporal stability of the Lake Baikal snow cover was controlled directly by using a device sensitive to intensity in the optical band (``luxmeter''). As well, Fig.~\ref{fig29} shows the relative spectral sensitivity of this luxmeter, peaked at about 550 nm and covering the wavelength range from 400 to 700 nm. Multiple measurements of intensity performed with this device pointed towards the snow cover at the normal incidence yielded almost the same results with the total relative standard deviation less than 4 \%. Likewise, the typical difference between intensity values measured at different locations was found to be of the order of 5 \% or less. Thus, we conclude that the albedo of the Lake Baikal snow cover is high ($K\approx$0.9), well defined (with a typical standard deviation $\delta K\approx$0.05 or less) and has good temporal and spatial stability. Remarkably, the standard deviation of the typical snow optical properties appears to be lower than the standard deviation of the typical atmospheric optical properties. The last nuisance factor is not unique for the SPHERE experiment and is common for all EAS Cherenkov detectors and atmospheric Cherenkov telescopes, such as H.E.S.S.~\cite{hin04, bon17}, MAGIC~\cite{lor04, ale15}, and VERITAS~\cite{kre04, par15}. Finally, we note that the energy scale uncertainty introduced by the variations of snow albedo is again of the order of several percent, which is typical for EAS experiments.

\subsection{Bidirectional reflectance of snow \label{ssect:snow-brdf}}

\begin{figure*}[t]
\begin{minipage}[t]{0.47\textwidth}
\includegraphics[width=19pc]{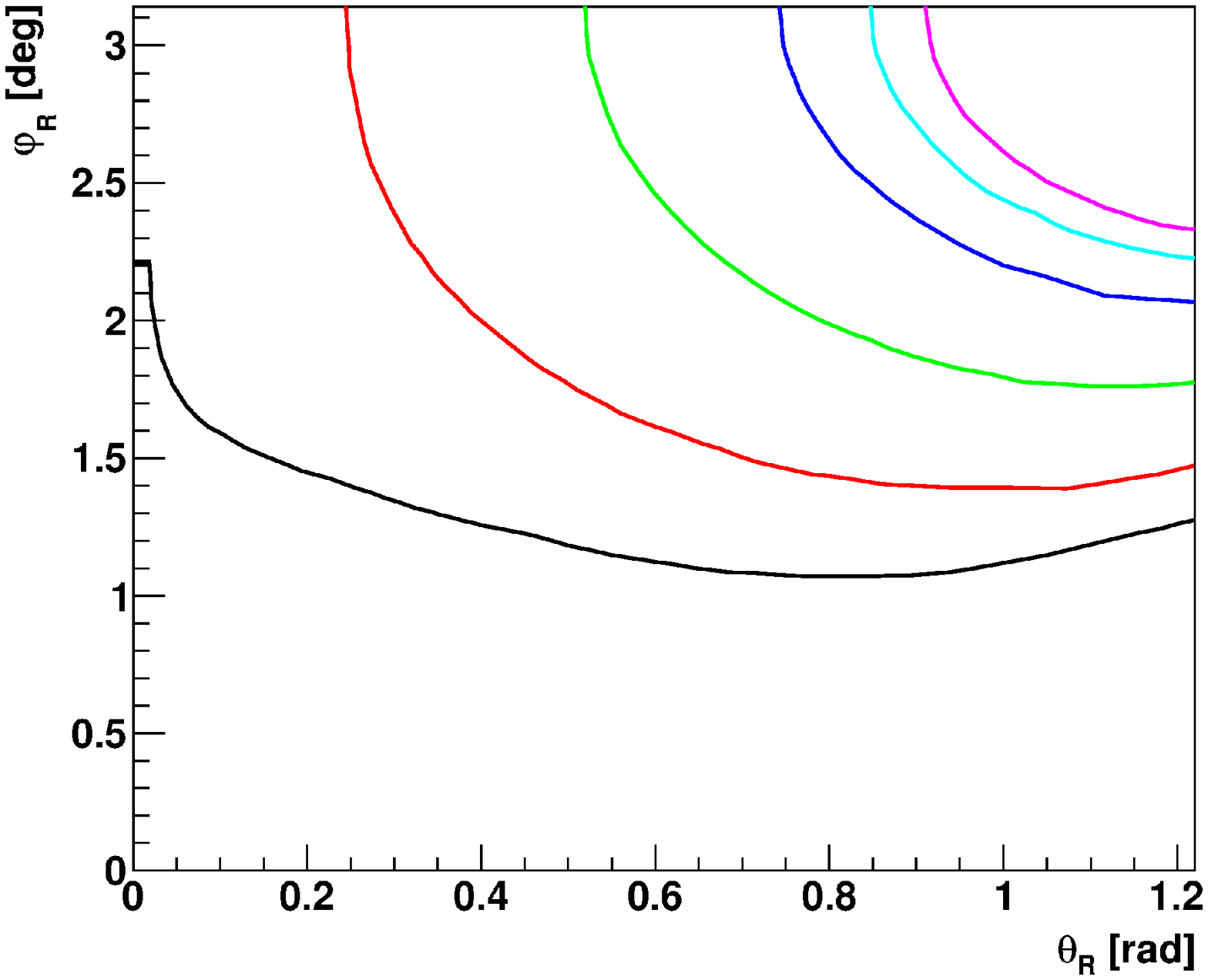}\hspace{1pc}%
\centering
\caption{BRDF contours in the model of \cite{li07}. Different values are denoted by various colours: black --- 0.74, red --- 0.76, green --- 0.80, blue --- 0.85, cyan --- 0.88, magenta --- 0.90.}
\label{fig30}
\end{minipage}
\hfill
\begin{minipage}[t]{0.47\textwidth}
\centering
\includegraphics[width=19pc]{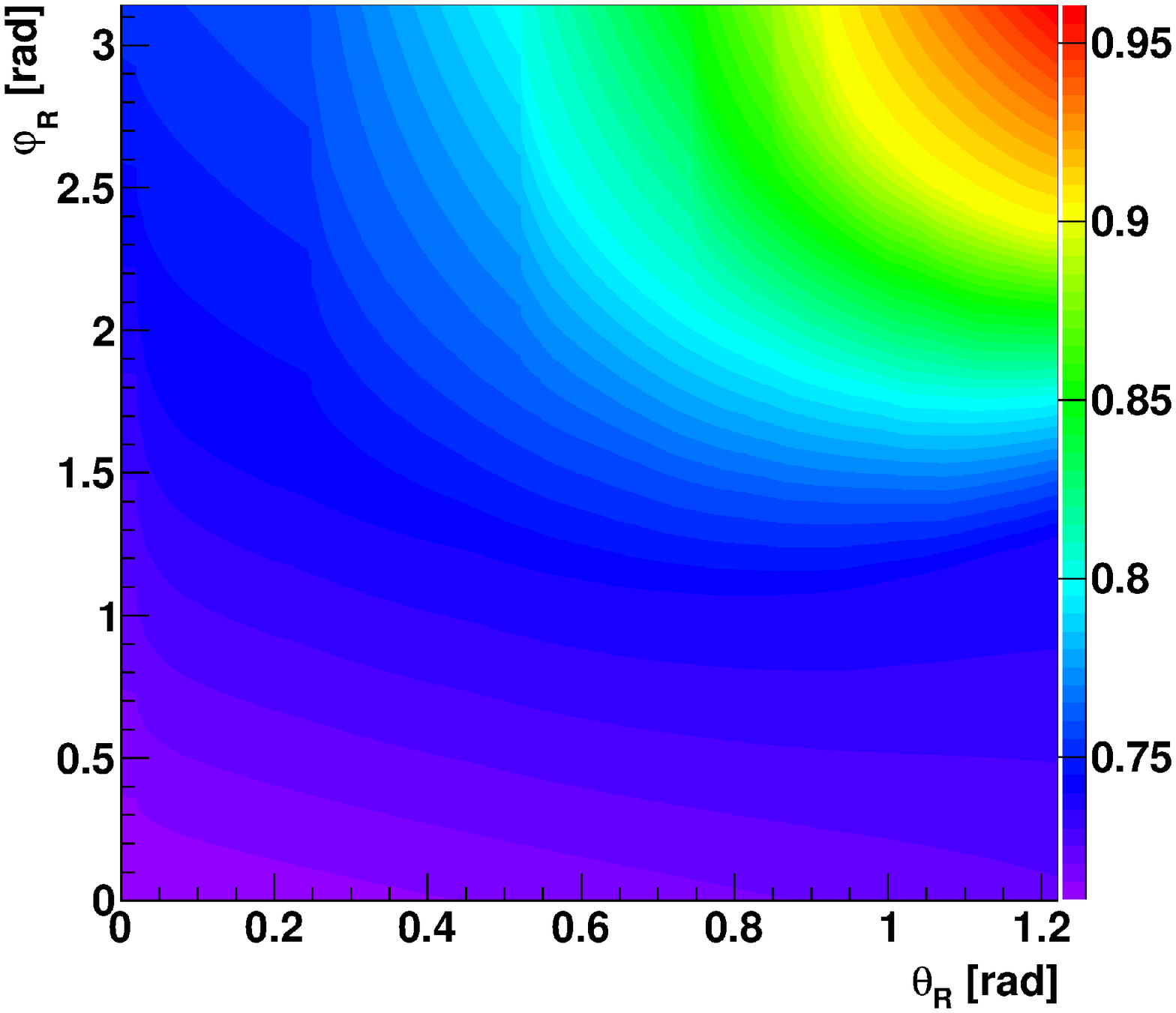}\hspace{1pc}%
\caption{Model of the BRDF used in this work. Color denotes the BRDF value vs. the reflection angles.}
\label{fig31}
\end{minipage}
\end{figure*}

The radiation reflected from a snow cover may have a complex distribution over the reflection angles. This distribution is characterized by the so-called ``bidirectional reflectance distribution function'' (BRDF)~\cite{war82}. BRDF may also strongly depend on the wavelength. A detailed study of the BRDF usually requires special hardware and significant investment of time (e.g.\cite{dum10}). We had performed a measurement of a simple proxy of the BRDF, namely, the dependence of the reflected intensity on the reflection zenith angle $\theta_{R}$ while all other conditions were fixed (see Fig.~11 in~\cite{ant15a}. These measurements were conducted at night using a light source shining at the normal incidence to the snow cover. The primary light source is LED; its radiation is reradiated by a luminescent layer. Therefore, the emission spectrum of this combined light source has two peaks around 450 nm and 550 nm (see blue curve in Fig.~\ref{fig29}). The same luxmeter that was already described in the previous Subsection was utilised as a photometric device. It was found that the relative photometric brightness $B_{r}(\theta_{R})= B(\theta_{R})/B(\theta_{R}=0)$ (i.e. $B$ normalized to unity at $\theta_{R}$= 0) reconstructed from these measurements is weakly dependent on $\theta_{R}$, namely, 0.93$<B_{r}(\theta_{R})<$1 for 0$<\theta_{R}<$40$^{\circ}$. Thus, in this case the reflection is almost isotropic.

Nevertheless, for the more general case of a non-normal incidence the anisotropy of the reflected radiation could be higher and, moreover, not axially-symmetric. Therefore, we have performed a detailed study of the impact that a modification of the BRDF has on the distribution of the LDF steepness parameter $\eta$. We compare simulated distributions of $\eta$ for two options of the BRDF: 1) BRDF= Const. for all values of the reflection angles (this option corresponds to the case of a Lambertian reflection) 2) a more realistic option for the BRDF following \cite{li07}. Fig.~\ref{fig30} shows several contours of the BRDF according to~\cite{li07} vs. reflection angles for the case of the incidence angle $\theta_{0}$=45$^{\circ}$ and $\lambda$= 412 nm. We note that the BRDF is practically axially symmetric (i.e. it is weakly dependent on $\phi_{R}$) for the case of small $\theta_{R}$ (below 0.2-0.3 rad) but is much more axially asymmetric at greater values of $\theta_{R}>$0.7-0.8 rad. In order to apply this model, we need to evaluate the BRDF between these contours. Fig.~\ref{fig31} shows an interpolation of the dataset presented in Fig.~\ref{fig30} to the full range of the reflection angles under our consideration: $\theta_{R}$ from 0 to 1 rad and $\phi_{R}$ from 0 to 2$\pi$. Fig.~\ref{fig31} demonstrates that the main features of the BRDF plotted in Fig.~\ref{fig30} are correctly reproduced by our interpolation. Although we have used a simplified linear interpolation technique and, as a result, the interpolated BRDF is not smooth, this interpolation is still sufficient for us, because the impact of the BRDF change will be shown to be small.

We simulated four sets of model detector response events for the case of zenith angles from 0 to 20$^{\circ}$ drawn from the isotropic distribution and observation altitude $H$= 400 m. The first and second sets both represent the case of primary protons with $E$=10 PeV but with different options for the BRDF (see above). The third and fourth sets are analogous with the first two ones, but these were calculated for the case of primary Iron. The procedure for simulating model response events was aready covered in details in Sect.~\ref{sect:response}. While calculating response events for the case of the modified BRDF we have utilised the same arrays of photoelectrons (see Subsect.~\ref{ssect:response-optics}) as for the case of the Lambertian reflection, but the contribution of every photoelectron to the model response event array (see Subsect.~\ref{ssect:response-electronics} and Fig.~\ref{fig14}--\ref{fig17}) was weighted according to the intensity change introduced by the modified version of the BRDF. Even though primary zenith angles of EAS under our consideration are limited to 20$^{\circ}$, for the case of the modified BRDF (the second option) we use the one presented in Fig.~\ref{fig31} with the incidence angle of 45$^{\circ}$. This choice is conservative because the anisotropy factor of the reflected radiation as well as the axial asymmetry, as a rule, falls with the decreasing incidence angle \cite{dum10}. Therefore, in reality the difference between the distributions of $\eta$ would be even smaller than for the case under consideration. We did not add any additional noise or background in this simulation in order to highlight the difference introduced by the BRDF change more clearly, without any unnecessary contaminating factors. The fluctuations of the model signal, however, were fully accounted for, as was described in Sect.~\ref{sect:response}. 

Using the arrays of model response events, for every event we calculate LDF by direct summation over time bins. We note that this simplified procedure may introduce some methodical uncertainty to LDFs obtained in this way, and thus somewhat decrease the separability of the classes of primary nuclei. A more sophisticated approach would include an account of the pulse shape in every measurement channel, for instance by a direct approximation of these pulses with subsequent estimation of the integral intensity in the measurement channels. This new procedure is now under development in our group and will be reported in a separate publication. 

Now we are in position to calculate the LDF steepness parameter $\eta$ for every event. The $\eta= \eta(0,70,70,140)$ parameter is defined as the ratio of the number of photons in the circle with the radius of 70 m to the same quantity in the ring with the inner radius of 70 m and the outer radius of 140 m. Both circle and ring have the same center coincident with the axis coordinates of the model shower. For every individual model LDF we introduce the same corrections as for the composite model LDFs, as described in Sect.~\ref{sect:composite}. After these corrections, the individual model LDFs become more axially symmetric and match better the corresponding composite model LDFs (see Fig.~\ref{fig6} and Sect.~\ref{sect:composite}). Then, we select the composite model LDF that fits the corresponding individual model LDF the best (i.e. has the minimal chi-square with respect to the individual model LDF) and compute the integrals over the above-indicated areas on the composite model LDF. The ratio of these integrals is our estimate of the $\eta$ parameter. This procedure is sensitive to Cherenkov light intensity in the central, sharply-peaked area of LDFs (see Fig. 1 in \cite{ant15c}).

\begin{figure}[tbh]
\begin{minipage}[h]{0.47\textwidth}
\centering
\includegraphics[width=19pc]{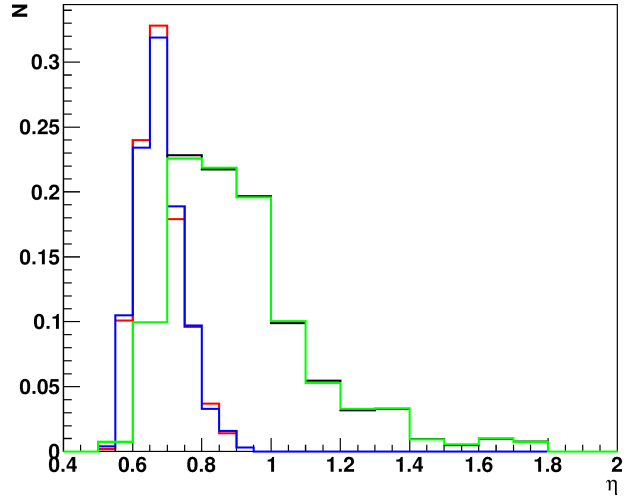}\hspace{1pc}%
\caption{Histograms of the LDF steepness parameter $\eta$. Black denotes protons and Lambertian reflection, red --- Iron and Lambertian reflection, green --- protons and modified BRDF, blue --- Iron and modified BRDF.}
\label{fig32}
\end{minipage}
\end{figure}

Finally, we compute histograms of $\eta$ for the above-defined four datasets, accepting only the events with the axis inside the FOV of the detector. These histograms are shown in Fig.~\ref{fig32}. We note that the separability of the nuclei classes is rather good; about 2/3 of protons with high values of $\eta$ could be selected with only 1.5--2 \% of contamination from Iron nuclei. Such steep-LDF proton showers develop significantly deeper in the atmosphere than most of Iron nuclei due to lower total interaction cross section of their primary particles. The histograms presented in Fig.~\ref{fig32} show only slight dependence on the assumed BRDF. Moreover, this dependence could be further suppressed by introducing an additional correction, as was done, for instance, in \cite{li07}. We leave this work for another paper.

\section{Performance for CR light nuclei spectrum measurement \label{sect:composition}}

Among many experimental problems of CR exploration, the task of inferring their mass composition from EAS parameters is particularly important and difficult (see e.g.~\cite{kam12} for a review). The classical approach to this problem in experiments with Cherenkov light usually utilizes the $X_{Max}$ parameter --- the depth of the shower maximum (e.g. \cite{fow01},\cite{iva09}). We note that the LDF steepness parameter allows to study the primary composition directly, {\it on the event-by-event basis}, without calculating any intermediate parameter such as $X_{Max}$ (see Fig.~\ref{fig32}). In this section we briefly consider the particular task of measuring the spectrum of CR light component to demonstrate the expected performance of the SPHERE experiment. The knowledge of this spectrum is important in context of models of the Galactic component of astrophysical neutrinos (e.g.,~\cite{ner16, guo14, ner18} that were registered by the Ice Cube observatory~\cite{aar13, aar13b, aar14, aar17}. Indeed, the energy of produced neutrinos strongly depends on the mass number of the primary nucleus of given energy. Moreover, this task have recently attracted significant attention of experimental groups~\cite{ape11, ape13, ape14}. Assuming that the observation time with the SPHERE-2 detector amounts to 300 hours, in Subsect.~\ref{ssect:composition-spec} we evaluate the expected relative uncertainty of the measured intensity of the light nuclei spectrum. In Subsect.~\ref{ssect:composition-exp} we estimate the total number of observational runs needed to attain the observation time of 300 hours for various experimental sites. For the rest of this section, we assume that the energy scale is known with precision $\sim$5 \%. Such absolute normalization may be obtained using results on the all-nuclei spectrum measured with EAS charged particles such as \cite{icetop13} (see also \cite{hor03} for discussion of energy scale spread in EAS experiments).

\subsection{Uncertainty of the spectrum of light nuclei \label{ssect:composition-spec}}

In what follows we assume that the primary CR differential all-nuclei spectrum is
\begin{equation}
J\approx 3\cdot10^{-9}\cdot\left(\frac{E}{10 PeV}\right)^{-3.0} \left[\frac{1}{PeV\cdot m^{2}\cdot s \cdot sr}\right] \label{eqn11},
\end{equation}
where $E$ is the primary energy measured in PeV. This estimate is not far from the results of~\cite{ape12}. In the 3--20 PeV energy region the spectrum is somewhat steeper (power-law index about 3.2), which would give some additional events below 20~PeV with respect to equation~(\ref{eqn11}). Other experiments, such as \cite{ame08, icetop13, abb18} give similar results on the all-nuclei spectral shape, but somewhat different results on the total normalization. The number of events inside the energy bin with the central energy $E$= 10 PeV and the total width of 3 PeV (which is roughly correspondent to 10 bins per decade of energy) thus will be $N\approx J\cdot(E_{Max}-E_{Min})\cdot A(E) \cdot T$, where $E_{Min}$= 8.5 PeV, $E_{Max}$= 11.5 PeV, $A$ is the acceptance, and $T\approx10^{6}$ s (approximately 300 hours) is the observation time. Assuming the observation altitude $H$= 400 m and $A\approx10^{5} m^{2}\cdot sr$ (which corresponds to 0-20$^{\circ}$ zenith angle range), we get $N\approx 10^{3}$. 

In \cite{ant15c} we show that about 40 \% of protons could be selected vs. Nitrogen nuclei at 10 PeV, 0-20$^{\circ}$ zenith angle range, 400 m observation altitude (with 1 \% contamination of Nitrogen). This quantity is representative for the selection performance of the light component. Indeed, for this case we have effectively the following four classification tasks: \\
1) selection of protons vs. the background of Iron nuclei (this is the easiest task of the four; about 90 \% of protons may be selected with 1 \% contamination of Iron nuclei \cite{ant15c}) \\
2) selection of protons vs. the background of Nitrogen nuclei (the performance of the proton selection is about 40~\%, as indicated above) \\
3) selection of Helium vs. the background of Iron nuclei (about the same performance as for the proton-Nitrogen separation task) \\
4) selection of Helium vs. the background of Nitrogen nuclei (this is the hardest task of the four). \\
In average, we may assume that the performance of selection of light nuclei averaged over these four cases is the one for the task of the proton-Nitrogen separation.

As an example, we assume the model of primary composition of~\cite{gai12}. At 10 PeV it predicts about 15 \% of protons and about 35 \% of Helium nuclei in the total flux, in total about 50 \% for the light component, which corresponds to about 500 events for the 8.5--11.5 PeV energy bin considered above. Out of these, about $N_{Sel}$=200 events would be identified as light nuclei. Thus, the relative statistical uncertainty would be $\epsilon_{Stat}\approx 1/\sqrt{N_{sel}} \approx 0.07$. The dominant systematic uncertainties are: \\
1) The spread of the primary energy for the light component nuclei vs. the primary composition $\epsilon_{E-Syst}\approx$ 1-3 \% \cite{ant15c} (see below for more details). $\epsilon_{E-Syst}$ is effectively the relative difference of mean reconstructed energy for proton and Helium primaries. We assume $\epsilon_{E-Syst}$=2~\%. \\
2) The spread of the acceptance for the light component nuclei vs. the primary composition $\epsilon_{A-Syst}\approx$ 3--6 \% (see Figs.~\ref{fig18}--\ref{fig19}). We assume $\epsilon_{A-Syst}$=5 \%. \\
3) The contamination of incorrectly classified heavy nuclei $\epsilon_{Cont}\approx$ 1 \%. \\
Thus, the total relative uncertainty 
\begin{equation}
\epsilon_{Tot}\approx \sqrt{\epsilon_{Stat}^{2}+\epsilon_{E-Syst}^{2}+\epsilon_{A-Syst}^{2}+\epsilon_{Cont}^{2}}\approx 0.09 \label{eqn12}.
\end{equation}

Here we neglected the migration of events between energy bins (this effect is important only at the highest energies) and the energy scale uncertainty. We also assumed the energy reconstruction approach of \cite{ded04,ant15c,ant15a}. Namely, we normalize the LDF of an observed EAS (with the unknown primary energy denoted as $E_{Exp}$) to composite model LDFs for a selection of model events. The normalization factor for the best-fit composite model LDF (with the known primary energy $E_{MC}$) is $K_{E} \approx E_{Exp}/E_{MC}$, therefore $E_{Exp} \approx K_{E}\cdot E_{MC}$. This approach utilizes information both on the experimental LDF normalization and shape, thus allowing for a significant reduction of the dependence of average estimated $E_{Exp}$ on the primary nucleus mass. In particular, the difference of estimated $<E_{Exp}>$ for proton and Iron is typically in the range of 1--6 \% (see Fig. 3 of \cite{ant15c}). Therefore, for the case of light nuclei only it is a fair assumption that $\epsilon_{E-Syst}\approx$ 1-3 \%. The typical energy resolution (statistical energy uncertainty) is also shown in Fig. 3 of \cite{ant15c} and ranges from 11 \% to 22 \% depending on the observation altitude, primary energy and primary nucleus mass.

The LDF steepness parameter $\eta$ is the main parameter sensitive to the primary nucleus mass in our approach. $\eta$ weakly depends on the primary energy, revealing a shift $\sim$0.1--0.2 per decade of energy, depending on the primary zenith angle value. Assuming the energy scale uncertainty of 5 \%, we estimate the additional (with respect to eq.~\ref{eqn12}) systematic uncertainty arising from the $\eta(E)$ dependence to be $\sim1-2$ \%.

The number of Cherenkov light photons arriving at the snow level depends on atmospheric conditions (e.g. \cite{ber00}). The level of atmospheric extinction (i.e. the decrease of the number of Cherenkov light photons arriving at the ground level) for the same geographical region as in the SPHERE experiment (namely, for the conditions of the Yakutsk experiment) was estimated in \cite{gri12} to be $\approx$30 \%. The ground level of Yakutsk is about 100 m a. s. l., while for the case of Lake Baikal, it will be remembered, the snow level is 455 m a. s. l. We note that the simulation setup of \cite{gri12} is very similar to the one of the present paper.

A detailed discussion of atmospheric conditions in context of Cherenkov light observations is available in \cite{ber00}. Atmospheric effects on the energy scale and the LDF steepness may be partially compensated if we know the absolute normalization of the spectrum, as was assumed above. Indeed,~\cite{ber00} shows that the impact of atmospheric effects on these quantities is not independent; namely, the change of the LDF steepness (this quantity is correlated with the $X_{Max}$ parameter) is usually accompanied by the corresponding change of the absolute normalization of the LDF.

Finally, we reconstructed pedestals in experimental event frames with EAS signal in the frame and found that the typical fluctuations and variations of these pedestals during the duration of the event frames (before and after EAS pulses) are $\sim$0.2 code units, what is much smaller than the typical fluctuation of the recorded EAS signal. Therefore, noise that is present in the event frames does not introduce an appreciable additional systematics with respect to eq.~\ref{eqn12}.

We note that the statistical uncertainty $\epsilon_{Stat}$ is dominant; therefore the precision of the light nuclei spectrum measurement would improve with more statistics. We argue that such a measurement would significantly contribute to the field of astroparticle physics because the total uncertainty of the measured light nuclei spectrum is comparable or even smaller than the uncertainty of other experiments such as~\cite{ape14, icetop13a}. The result of the measurement would have some dependence on the model of high energy nucleus-nucleus interaction~\cite{ost06, Ahn09, epos09}. We also note that we do not deem it possible to outperform ground-based experiments observing the charged component of EAS in terms of the registered number of events. This is due to low duty cycle of the SPHERE experiment. This problem is common for Cherenkov experiments. 

\subsection{The observation time and the number of observation runs \label{ssect:composition-exp}}

Now we estimate the number of observation runs (assuming one run per year) sufficient to obtain the observation time of 300 hours. First of all, let us consider the case of the Lake Baikal snow cover. At Lake Baikal, ice typically sets in mid-January. However, in order to operate conveniently, we need rather thick ice cover (at least 10-15 cm) in order to be able to sustain the weight of heavy equipment needed for the launch of the BAPA balloon. Indeed, about 50 vessels with total mass about 3 tons are typically needed to fill the balloon with Helium gas. This circumstance delays the start of observations by about two weeks. Snow cover with sufficiently good optical properties is typically present at Lake Baikal until the end of March. Therefore, we typically expect that two moonless periods (namely, the ones in February and March) are available for work in this case. For the case of the 2018 conditions, we estimate the total duration of these moonless periods to be about 290 hours. Assuming that clouds, strong wind and other nuisances reduce this time by the factor of two, the total observation time for one year at Lake Baikal amounts to 150 hours. This is an optimistic estimate, because technical and financial problems may significantly reduce this effective time. In real conditions we were typically able to obtain 30-35 hours per year, mainly due to the shortage of money. However, our experience makes us believe that 100-150 hours of observation time per year are attainable. Therefore, the level of precision set in the previous Subsection is attainable in 2-3 years.  

For the case of observations in Antarctica, the total effective observation time per year may be much longer than for the case of Lake Baikal. Assuming that dark time amounts to 1/2 of year, the moonless periods --- 1/4 of year, and clouds, auroras or other nuisances reduce the last estimate by the factor of two, we estimate the total observation time to be about 1000 hours. Therefore, the level of precision set in the previous Subsection is attainable in only one observation run.  

\section{Discussion \label{sect:discussion}}

In the present paper we describe the expected spatial and temporal structure of reflected Cherenkov light signal as observed by the \mbox{SPHERE-2} detector. This is the first work to assess the concept of CR study with reflected Cherenkov light using direct detailed MC simulations. We have developed a highly modular code for detector response simulation, starting from the level of Cherenkov light properties on the ground, and permeating many stages of simulation such as parameters of individual photoelectrons, pulses in individual measurement channels, trigger response for individual events, including global quantities averaged over large samples of model events such as the instrumental acceptance. The quality of trigger response simulation is critical for correct reconstruction of the primary spectrum and composition~\cite{ant15a, ant15c}. Therefore, we have used amplitude thresholds taken directly from experimental data. These thresholds are recorded at the start of every measurement period and define the response of amplitude discriminators. We also have verified that experimental events classified as EAS indeed cause the triggering of our model trigger function. 

Such a multi--stage approach is especially helpful for the case of the reflected Cherenkov light method, since fluctuations of observable signal are large and observable parameters are strongly distorted by the detector. However, we expect that a similar analysis could be also useful for other experiments and projects, including the balloon--borne experiment JEM-EUSO-Balloon~\cite{adam15c}, the satellite project JEM-EUSO~\cite{adam15b}, as well as for the ground-based Yakutsk experiment~\cite{iva09} and the NICHE project~\cite{niche15a, niche15b, niche17}.

The reflected Cherenkov light technique has a number of similarities and differences with other methods developed for EAS observation. In terms of observable signal properties, the JEM-EUSO project is the most similar to the SPHERE experiment. However, there is a major difference between the corresponding detection methods. The JEM-EUSO instrument is designed to register mainly EAS fluorescent light, while the \mbox{SPHERE-2} telescope is aimed at the Cherenkov light observation. Another important consideration is that the typical projection area of one JEM-EUSO pixel to the ground surface would be $S_{p}\sim$ 1~km$^{2}$ \cite{adam15a}, while the same value for the \mbox{SPHERE-2} telescope is about three orders of magnitude smaller, allowing a detailed measurement of the LDF shape.

The approach to neutrino and CR detection using radio emission reflected from ice in the ANITA experiment~\cite{sch15, gor16a} is in many aspects similar to the observation of the reflected Cherenkov light. The main advantage of the radio method is its larger duty cycle ensuring a greater observation time, while EAS Cherenkov light detection with any Cherenkov telescope is only possible during clear moonless nights. Unfortunately, simulations of the observable radio signal properties are still of considerable difficulty~\cite{alv15, bel16a} and suffer from various systematic uncertainties (e.g.~\cite{cor17a}) that eventually transfer to significant systematic uncertainties in the all-nuclei spectrum and composition~\cite{tho16a, bui16a}.

A ground-based instrument similar to the \mbox{SPHERE-2} telescope with the optical axis pointed towards the horizon could also be useful for CR composition studies and even neutrino detection. Indeed, in \cite{ner16a} it was shown that Cherenkov light produced by muons of near-horizontal EAS can serve as a probe of the CR composition. If located on top of a hill or a mountain that could be found around Lake Baikal in abundance, a SPHERE-type telescope could observe upward-going showers from the charged-current interactions of tau neutrinos inside the lake or in adjacent medium (see e.g.~\cite{far02a, ner17a}). Returning to the balloon technique, the SPHERE-An\-tarc\-ti\-ca project is currently under development~\cite{ant17}. It will significantly increase the exposure (that is, effective area times observation time) by at least 3--4 orders of magnitude compared to the \mbox{SPHERE-2} detector capabilities. For the SPHERE-An\-tar\-cti\-ca case, however, one pixel projection area is again $S_{p}\sim$ 1~km$^{2}$, not allowing detailed measurement of LDF shape similarly to JEM-EUSO. Another promising option is the SPHERE-HD project, aimed at observation of reflected Cherenkov light from altitudes up to 3 km with a detector consisting of several thousands or even several dosens of thousands pixels~\cite{ant15a}. It would allow to raise observation altitude, and thus the number of registered EAS, without an appreciable loss of data quality.

\section{Conclusions \label{sect:conclusions}}

We have discussed the method of extensive air shower observations by means of Cherenkov light, reflected from snow surface for the case of the \mbox{SPHERE-2} detector. Direct detailed MC simulations of detector response and instrumental acceptance show that this method allows for a detailed study of cosmic rays with energies above 10~PeV. We have introduced the concept of ``composite model quantities'', that allows to understand effects imprinted by the detector's non-ideality to lateral distribution function, observable pulse shape, and trigger response. We have demonstrated that the uncertainty of the snow cover optical properties does not introduce an appreciable systematic uncertainty of the lateral distribution function steepness distribution. Finally, we have shown that under favourable conditions the expected exposure is sufficient to allow the measurement of the spectrum of light CR nuclei with the total uncertainty comparable with other contemporary experiments. Results presented here will be used in subsequent publications dealing with all-nuclei spectrum reconstruction and composition study using the \mbox{SPHERE-2} telescope data.

\section*{Acknowledgements}

We are grateful to Prof. L.G. Dedenko and to E.V. Khalikov for helpful discussions. The work of T.D. was supported by the Munich Institute for Astro- and Particle Physics (MIAPP) of the DFG cluster of excellence ``Origin and Structure of the Universe'' at the late stage.

\section*{References}
\bibliography{Sphere-Signal.bib}
\end{document}